\documentclass[preprint2]{aastex}

 \usepackage{graphicx}

\newcommand {\ha} {H$\alpha$}
\newcommand {\hi} {\mbox{H{\sc i}}\,\,}

\newcommand {\kms} {\,km\,s$^{-1}$\,}
\newcommand {\M} {\mbox{${\cal M}$}}

\newcommand {\msol} {\M$_\odot$\,}

 \newcommand {\rml}{$\big({\cal M}/L_B\big)_\ast$ }

\slugcomment{Accepted for publication in The Astronomical Journal}

\shorttitle{\hi studies of NGC 24 and NGC 45}
\shortauthors{Chemin, Carignan, Drouin \& Freeman}

\begin{document}

\title{\hi studies of the Sculptor group galaxies. \\
VIII. The background galaxies: NGC 24 and NGC 45}
\author{Laurent Chemin\altaffilmark{1,2}, Claude Carignan\altaffilmark{1}, 
Nathalie Drouin\altaffilmark{1}, and Kenneth C. Freeman\altaffilmark{3}}

\altaffiltext{1}{Laboratoire d'Astrophysique Exp\'erimentale (LAE),
    Observatoire du mont M\'egantic, and D\'epartement de physique,
    Universit\'e de Montr\'eal, C.P. 6128, Succ. Centre-Ville, 
    Montr\'eal, Qc, Canada H3C 3J7}
 \email{[chemin,carignan]@astro.umontreal.ca, kcf@mso.anu.edu.au}
\altaffiltext{2}{Observatoire de Paris, section Meudon, GEPI, CNRS-UMR 8111 \& Universit\'e Paris 7, 5 Pl. Janssen, 92195, Meudon, France}
\altaffiltext{3}{Research School of Astronomy and Astrophysics, Mount Stromlo
Observatory, Weston Creek, ACT 2611, Australia}
 
\begin{abstract}
In order to complete our \hi survey of galaxies in the 
Sculptor group area, VLA observations of NGC 24 and NGC 45 are presented. 
These two galaxies of similar magnitude $M_B \sim -17.4$
lie in the background of the Sculptor group
and are low surface brightness  galaxies, especially NGC 45. 
The \hi distribution and kinematics are regular for NGC 24 
while NGC 45 exhibits a kinematical twist of its major axis. A tilted-ring model
shows that the position angle of the major axis changes by $\sim 25\degr$.
A best-fit model of their mass distribution gives 
mass-to-light ratios for the stellar disk 
of 2.5 and 5.2 for NGC 24 and NGC 45  respectively. 
These values are higher than the ones expected from 
stellar population synthesis models. 
Despite the large dark matter contribution, the galaxy mass 
is still dominated by the stellar component in their very inner regions.
These high mass-to-light ratios are typical 
of what is seen in low surface brightness
galaxies and may indicate that, in those galaxies, disks are far from the
maximum disk case. The halo parameters derived from
the best-fit models are thus lower limits. 
\end{abstract}

\keywords{galaxies: halos --- galaxies: fundamental parameter (mass) --- 
galaxies: individual (NGC 24, NGC 45) --- galaxies: kinematics and dynamics
galaxies: structure}

\section{Introduction}
\label{sec:intro}
At the beginning of the 90's, a series of papers was started 
on the \hi properties of
the Sculptor group galaxies (Puche \& Carignan 1988; Carignan \& Puche 1990a,b; Puche, Carignan \& Bosma 1990;
Puche \& Carignan 1991; Puche, Carignan \& van Gorkom 1991; Puche, Carignan \& Wainscoat 1991).
Since the Sculptor galaxies are all late-type spirals with very little or
no bulge, they are ideal candidates for determining
the basic dark halo parameters (Carignan \& Freeman 1985). Their simple disk +
halo stucture makes it easier to identify the contribution of each
component to the rotation curve. Also, since the \hi distribution 
of late-type spirals is
usually extended (Huchtmeier, Seiradakis \& Materne 1980), it is possible to
measure their rotation curve out to many disk scale lengths, an essential
condition to tie down the halo parameters (Lake \& Feinswog 1989).


\begin{figure*}\centering
\includegraphics[width=\columnwidth]{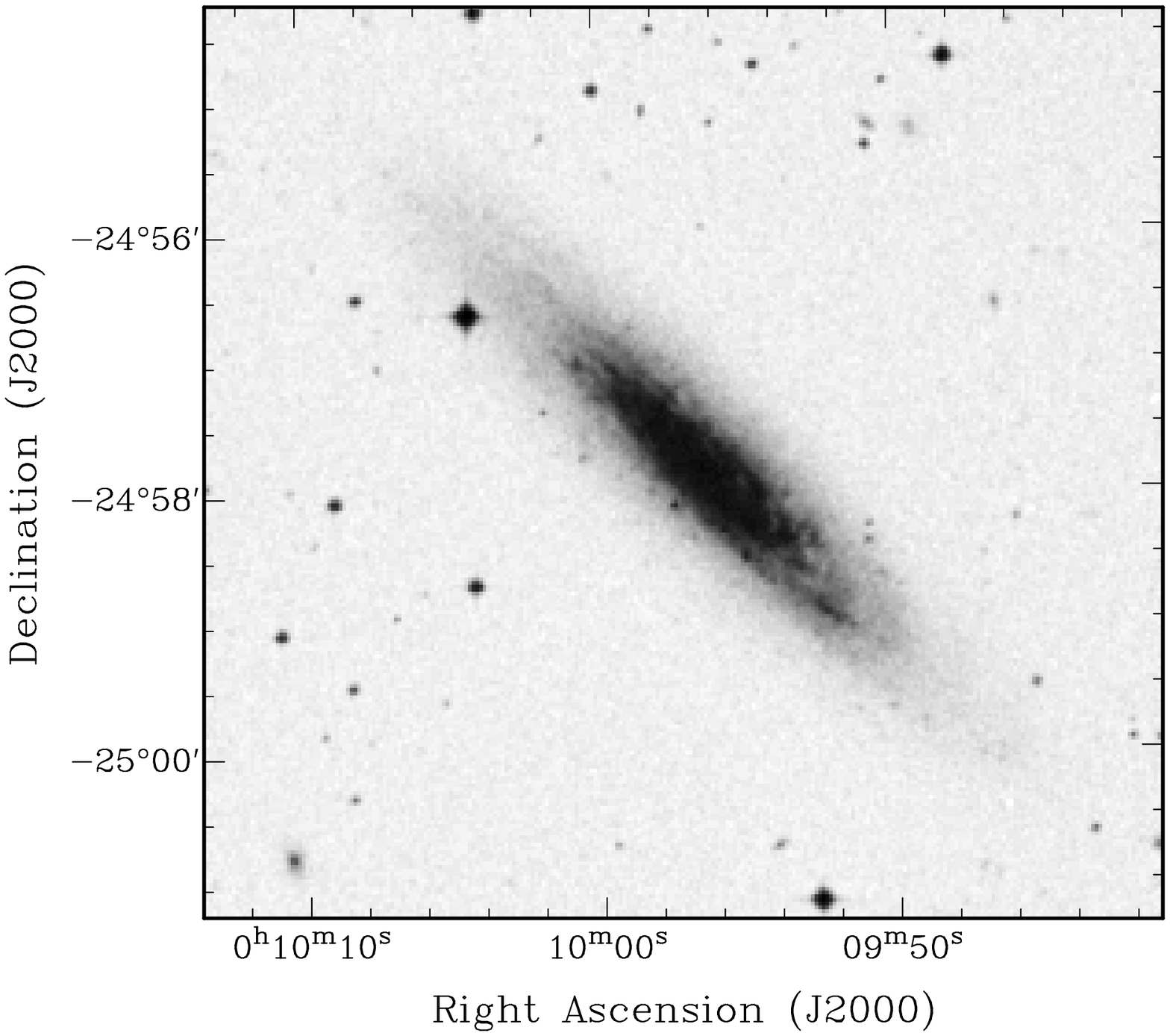}\includegraphics[width=\columnwidth]{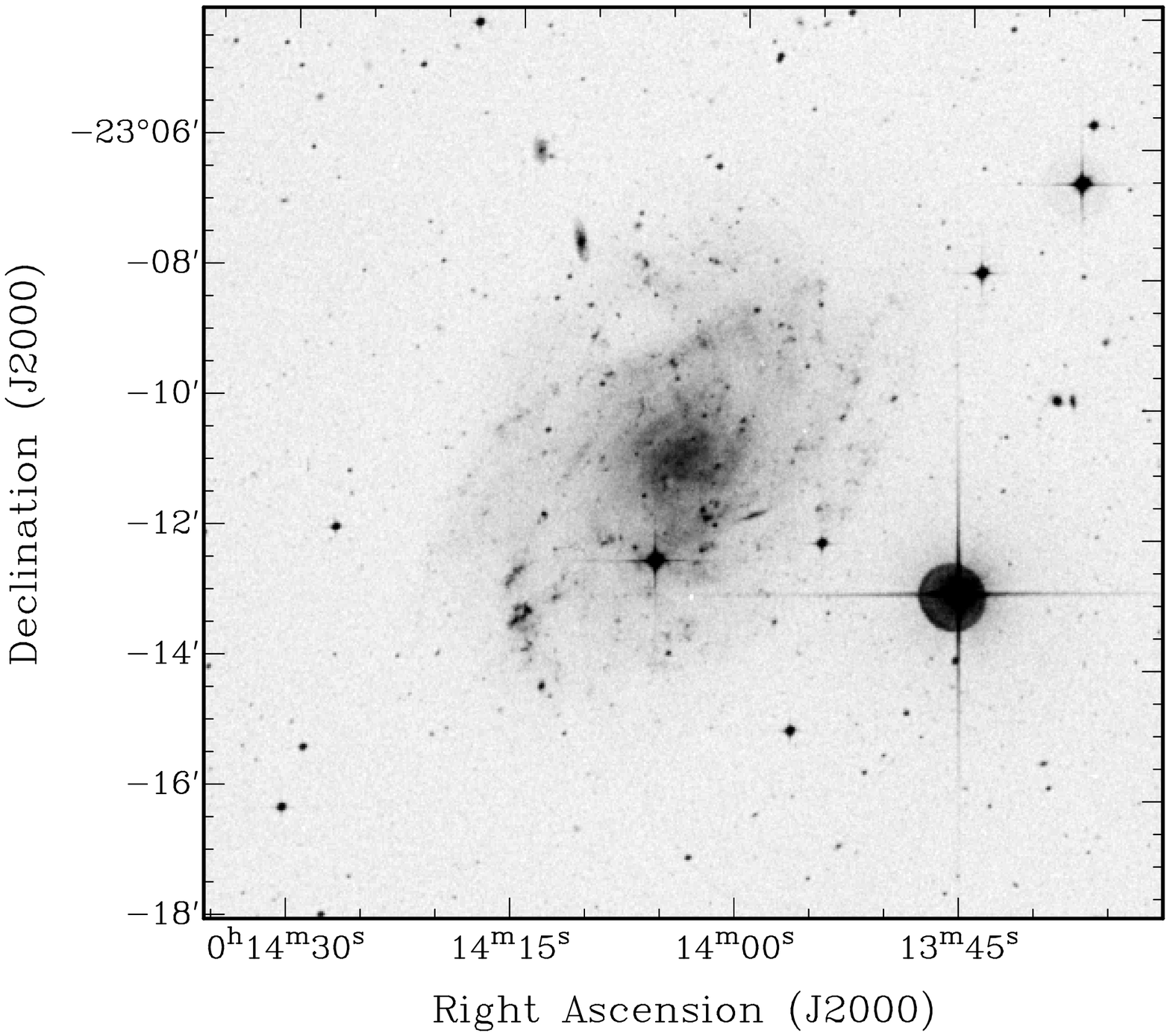}
\caption{DSS blue images of NGC 24 (left) and
NGC 45 (right). \label{dssimages}}
\end{figure*}

As described in Puche \& Carignan (1988), 
the Sculptor Group, the nearest group of galaxies from the Local Group, 
covers an area of $\sim 20\degr$ on the sky. In this area, besides the five main 
members studied so far (NGC 55, 247, 253, 300 and 7793), two more galaxies,
NGC 24 and NGC 45 (Figure~\ref{dssimages}), fall in the same region and have similar sizes and degree
of resolution as well as $\sim$ 20 known dwarf galaxies
(C\^ot\'e, Freeman \& Carignan 1997; C\^ot\'e, Carignan \& Freeman 2000). 
However, when looking at the estimated distances, a 
mean distance of $2.4 \pm 0.6$ Mpc is found for the main members 
(de Vaucouleurs 1978) but 6.8 Mpc (7$\sigma$) and
5.9 Mpc (6$\sigma$) for NGC 24 and NGC 45, respectively 
(based on a Hubble constant of 75 \kms Mpc$^{-1}$, Tully 1988). 
Moreover, when doing dynamical mass estimates of the group,
any combination of the main members gives consistently 
$\sim3 \times 10^{12}$ M$_{\odot}$ for the group, while including NGC 24 
and NGC 45 makes the estimate jump to $\sim3 \times10^{13}$ M$_{\odot}$,
clearly too large for a loose group like Sculptor.
All this suggests strongly that NGC 24 and NGC 45 are not members of the group 
but background objects.

Even if they were not members of the Sculptor group, it was decided
to study them because they belong to a very interesting subgroup of
galaxies called the low surface brightness (LSB) spiral galaxies, 
especially NGC 45.
The kinematical study of LSB galaxies currently feeds the ``cusp-core controversy'',  
(e.g. Swaters et al. 2003; Hayashi et al. 2004; de Blok 2005, and references therein).
One of the many important questions which is
brought to the forefront from the numerous studies of this type of
object and which motivates part of the present work, is whether or not LSBs
(and more generally low mass late-type spirals) can possibly form a
distinct class in the Hubble sequence. The
hypothesis emanating from studies on LSBs is: morphologically they belong
to the spiral class, but dynamically they look more like the
dwarf irregular class since they are dominated at all radii by their
dark halo (Jobin \& Carignan 1990; C\^ot\'e, Carignan \& Sancisi 1991; 
Martimbeau, Carignan \& Roy 1994). 

Moreover, a study of optical rotation curves of spiral galaxies
(Buchhorn 1992) showed convincingly that the optical rotation curves in the
inner parts of these galaxies give \rml ratios which are very high.
The most likely explanation for this effect
is that the dark matter component contributes a large fraction of the
gravitational field even in the inner regions of LSBs galaxies 
(de Blok \& Bosma 2002). 
Verheijen (1999) and Verheijen \& Tully (1999) have shown convincingly
the clear dichotomy between the rotation curves of HSB (high surface
brightness) galaxies, which are most of the time close to the
maximum disk situation, and those of LSB galaxies which are dominated at all radii
by the dark matter component. Thus, in terms of their mass distribution,
LSBs appear to be more closely related to dwarf irregular galaxies 
(e.g. DDO 154: Carignan \& Freeman 1988;
Carignan \& Beaulieu 1989) where the luminous component is known to make a
negligible contribution to the gravitational field everywhere in the
disk, than to massive spirals (e.g. NGC 6946:
Carignan et al 1990) where the luminous disk is the main contributor
in the inner regions and the dark component only contributes
significantly in the outer parts.


\begin{figure*}\centering
\includegraphics[width=\columnwidth]{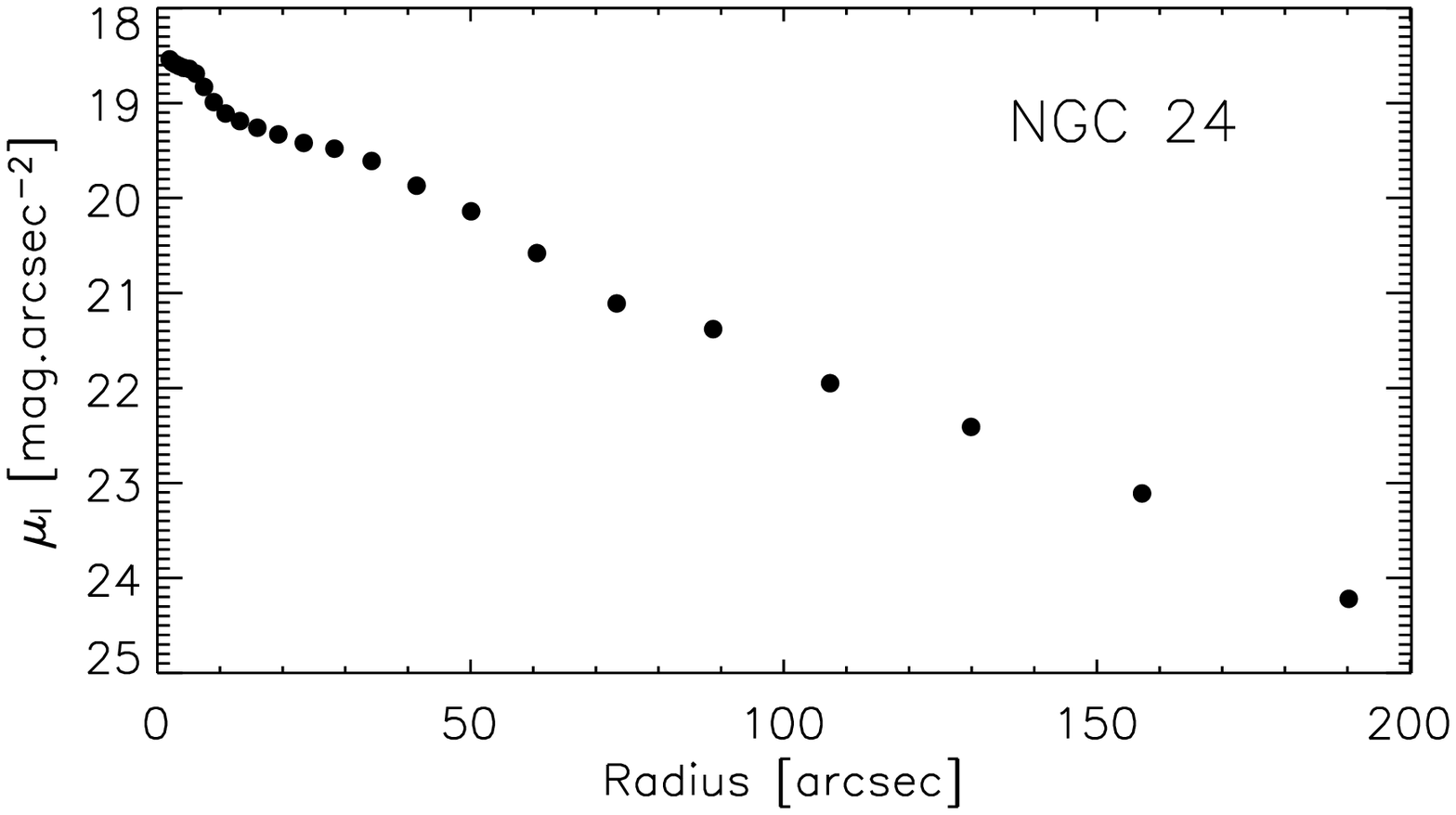}\includegraphics[width=\columnwidth]{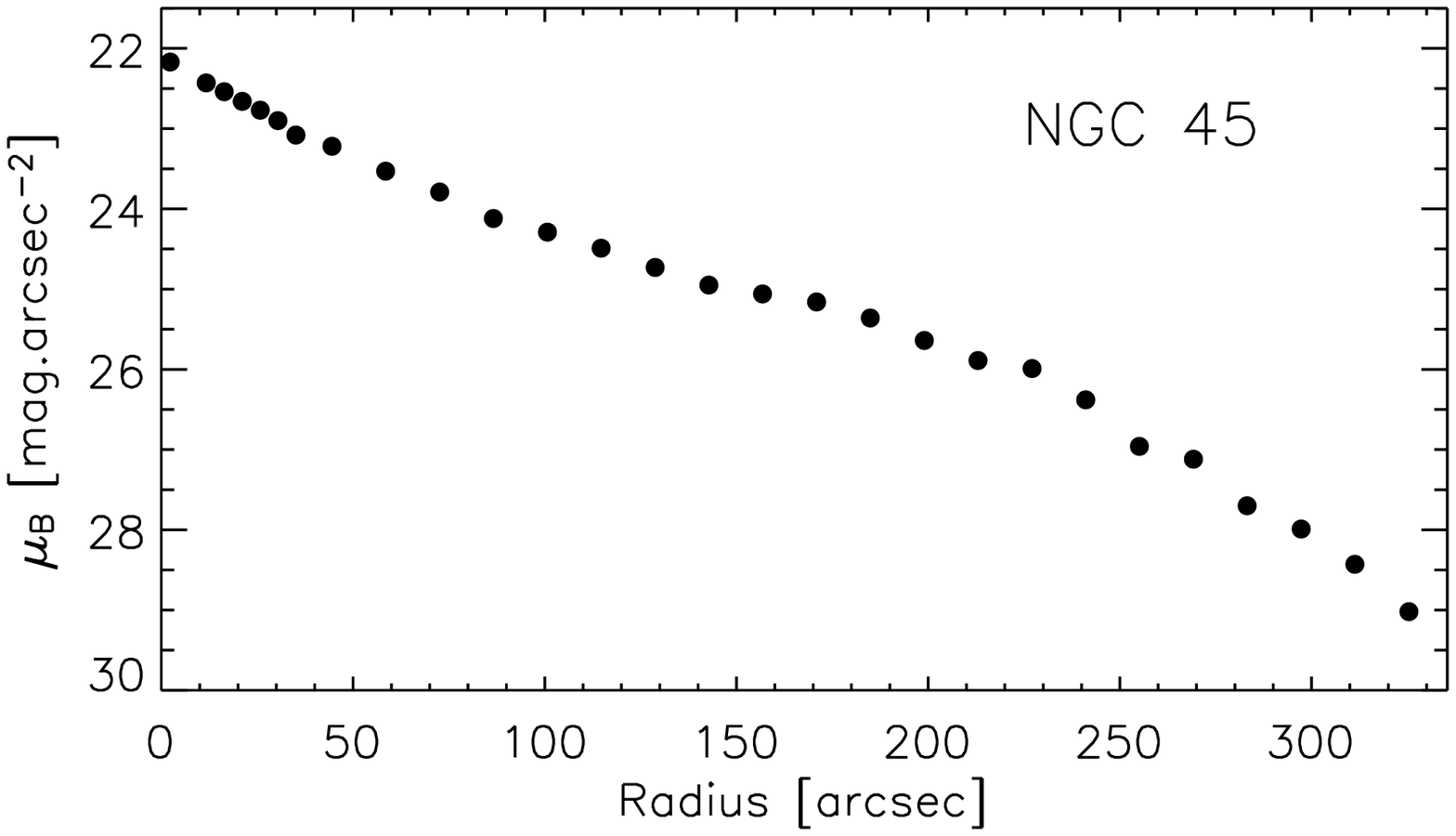}
\caption{$I-$band luminosity profiles of NGC 24 (left) and 
$B-$band luminosity profile (Romanishin et al. 1983) of NGC 45 (right).
\label{lp}}
\end{figure*}

In order to get a better understanding of the mass distribution in NGC 24 and NGC 45, both
optical and 21~cm radio observations were obtained and are presented
in Section~\ref{sec:obs}. In Section~\ref{sec:photometry}, the luminosity profiles and the optical properties
are discussed. After studying the \hi content and its
distribution in Section~\ref{sec:neutrgas}, Section~\ref{sec:kinematics} 
concentrates on the velocity field
of these systems and presents their rotation curves. 
In Section~\ref{sec:mass} the data are
analyzed in terms of a two component model: a luminous (stellar$+$gaseous) disk
and a dark halo represented by an isothermal sphere potential. Finally,
the discussion in Sec.~\ref{sec:discussion} is followed by a summary of the results in
Sec.~\ref{sec:conclusion}.  

\section{Observations}
\label{sec:obs}
\subsection{Optical data}
The surface photometry of NGC 24 was obtained from a set of
observations in the $I-$band using a 1024$\times$1024 Thompson CCD, 
with the f/1 focal reducer at the prime focus of the AAT 3.9 m. The
19\,${\rm \mu}$m pixels resulted in a resolution of 0.98$''$/pixel
for a total field of 16.7$'\times 16.7'$. The adopted mean extinction was
0.085 per airmass in the $I$-band. 
The data were obtained in 1990 September, with integration times of
20\,s and 5\,s. 
After the images were bias subtracted and flat-fielded using dark sky
flats, the foreground overexposed stars were removed. 
Pixels within a circular region were deleted and replaced
by a two dimensional surface function evaluated from the pixels lying
in a surrounding background annulus. Afterwards, the sky previously
evaluated by  averaging the signal of many regions depleted of stars was
subtracted from the galaxy image.


\begin{figure*}[!t]\centering
\includegraphics[width=\columnwidth]{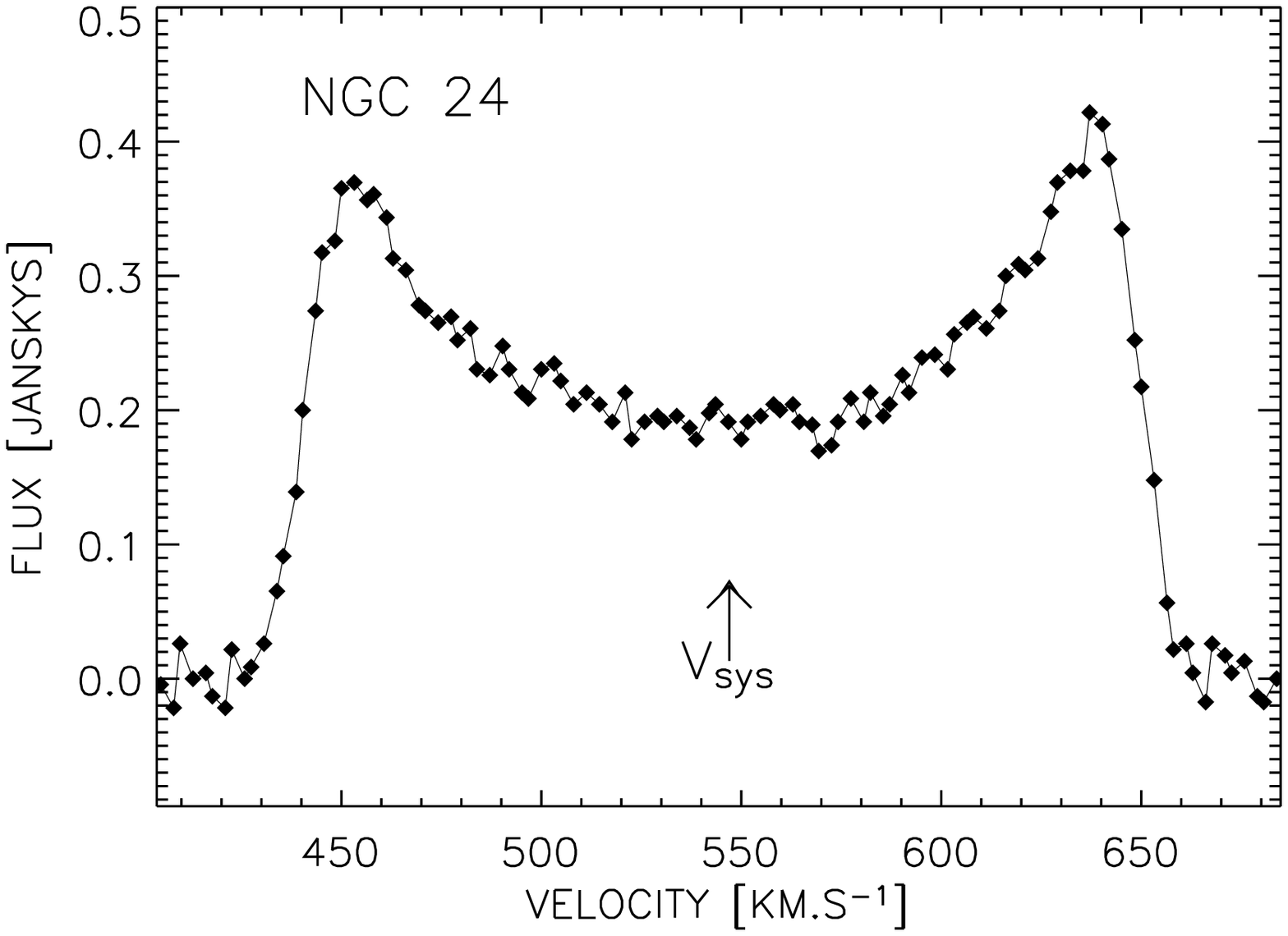}\includegraphics[width=\columnwidth]{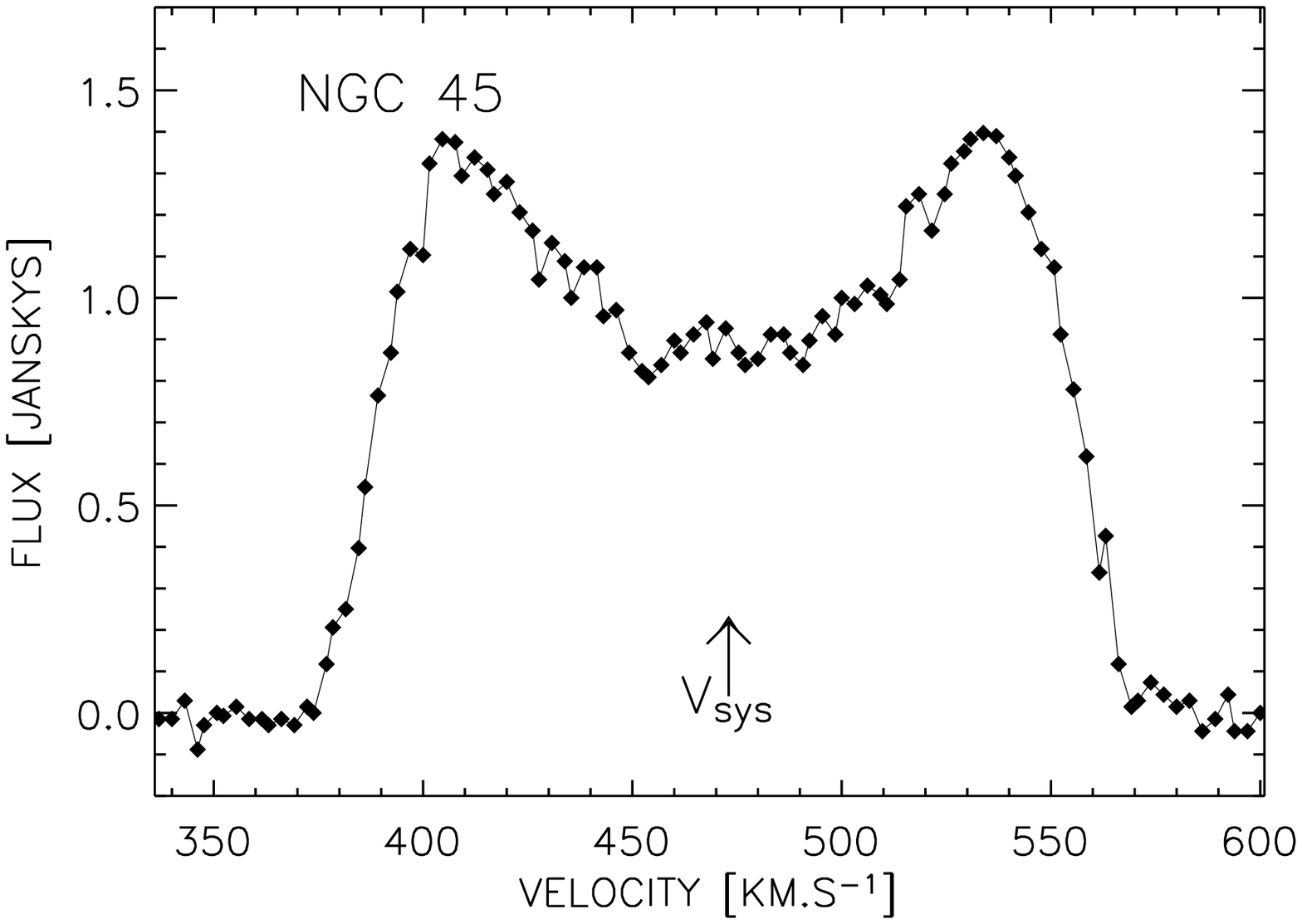}
\caption{\hi global profile for NGC 24 (left)
and NGC 45 (right) obtained by integrating the \hi
emission in each channel map. The arrows indicate the systemic velocity 
as computed from the profiles.}
\label{global}
\end{figure*}
 
\subsection{Radio data}

In order to get the kinematical information needed,
high sensitivity low resolution 
\hi line observations were obtained with the Very Large
Array (VLA\footnote{The NRAO is a facility of the National 
Science Foundation operated 
under cooperative agreement by Associated Universities, Inc}). 
They consist of 6\,h observations
for each galaxy (5\,h on source, 1\,h on calibrators), made in
1992 June. 
In order to get a better
{\it uv} plane coverage,
the hybrid DnC configuration, with larger antenna spacings
in the North Arm was selected,
since NGC 24 (${\delta}\,\simeq\,-25\degr$) and
NGC 45 (${\delta}\,\simeq\,-23\degr$) are at low declinations.
The candidates have been observed with a total
bandwidth of 1.56~MHz divided in 128 channels and using on-line Hanning
smoothing. This gives a channel separation of 12.2~kHz, 
corresponding to $2.5~{\rm km\,s^{-1}}$. The
parameters of the synthesis observations can be found in 
Tables~\ref{vla-n24} and~\ref{vla-n45}. 
No attempt to correct for the beam-smearing effect is done in this article.

\begin{figure*}\centering
\includegraphics[width=\columnwidth]{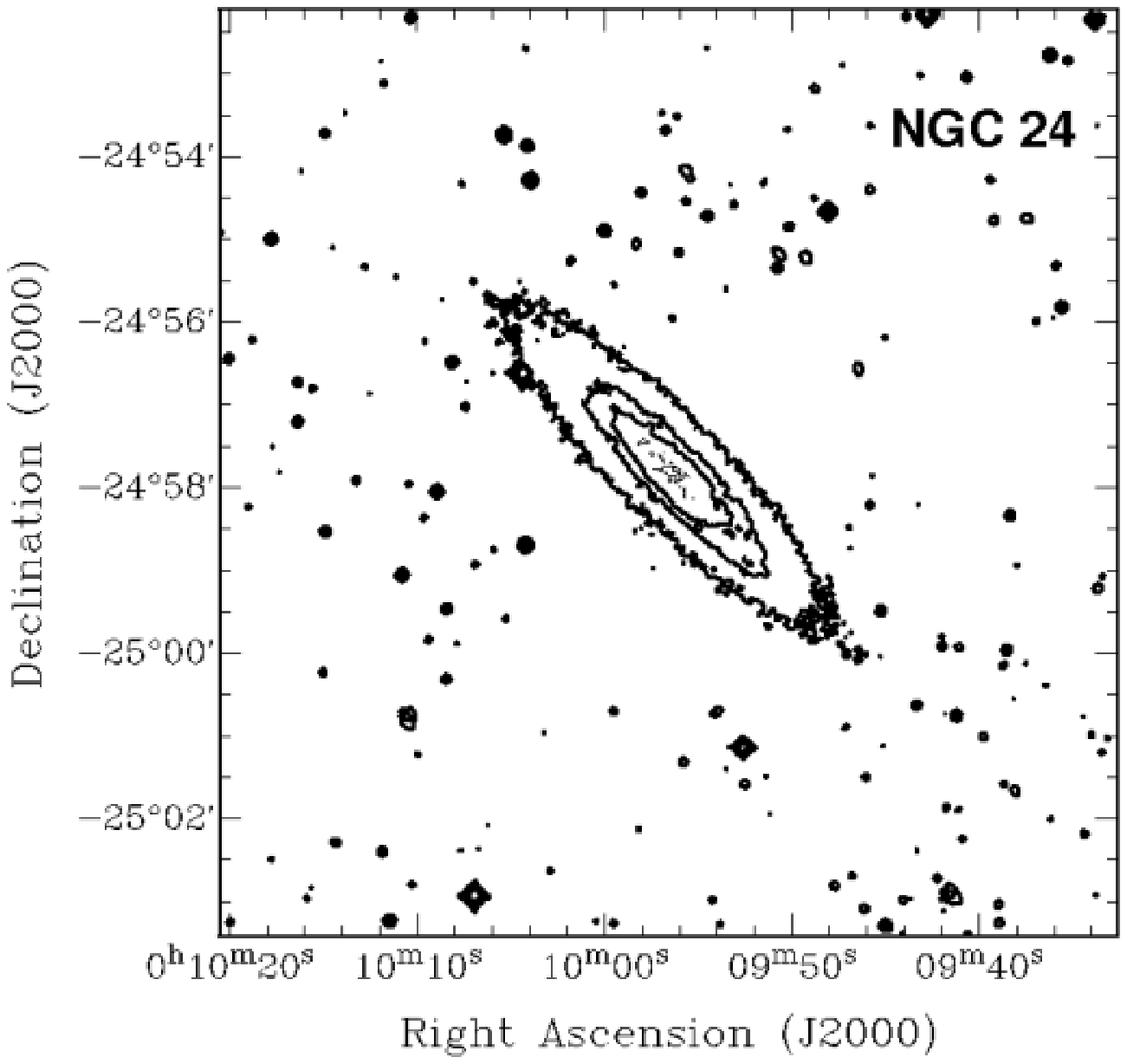}\includegraphics[width=\columnwidth]{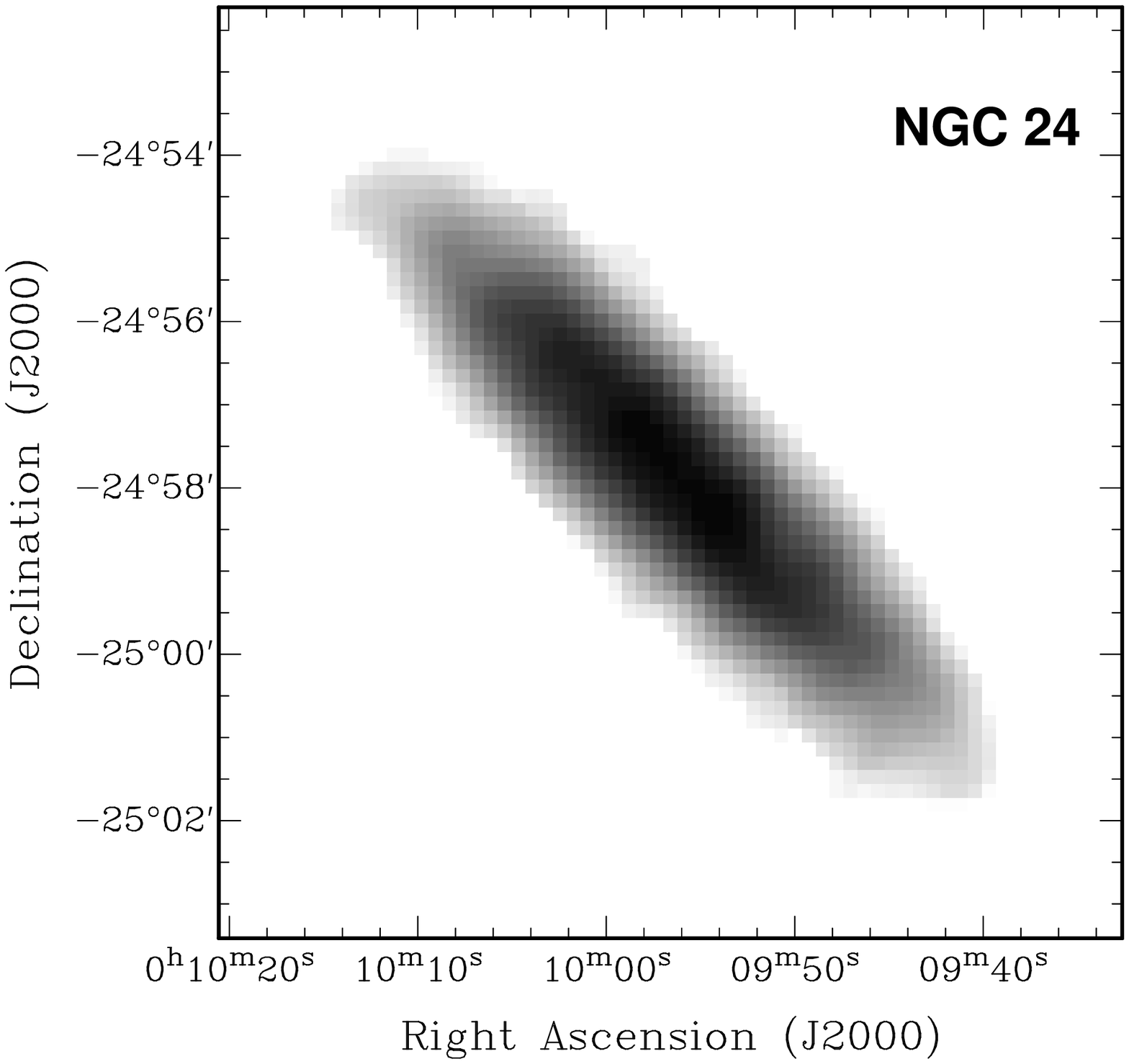}
\caption{NGC 24 : Contours of the DSS optical image (left). Greyscale total \hi map displayed using a logarithmic stretch (right).}
\label{contoptgreyhi_n24}
\end{figure*}

\begin{figure*}\centering
\includegraphics[width=\columnwidth]{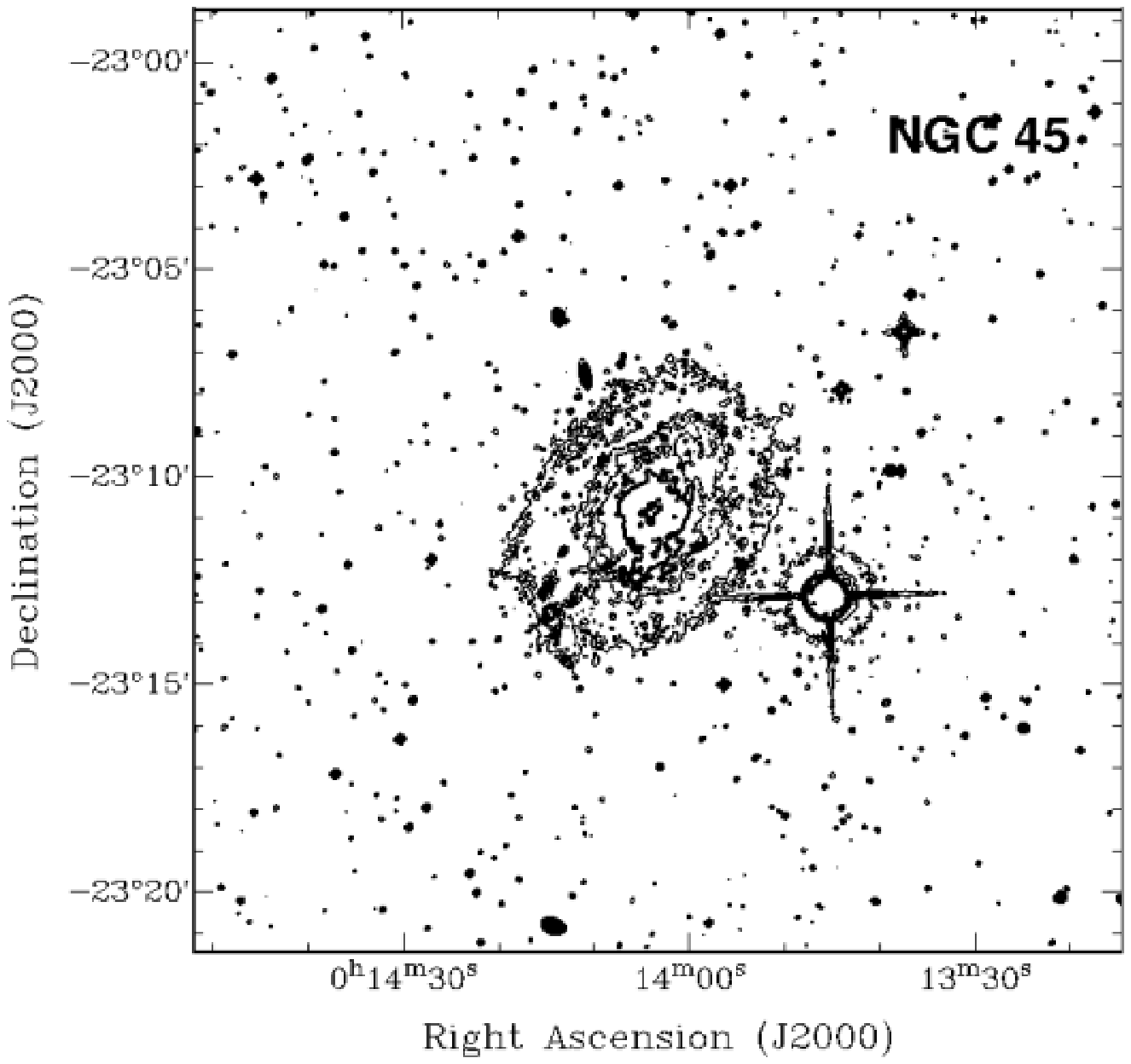}\includegraphics[width=\columnwidth]{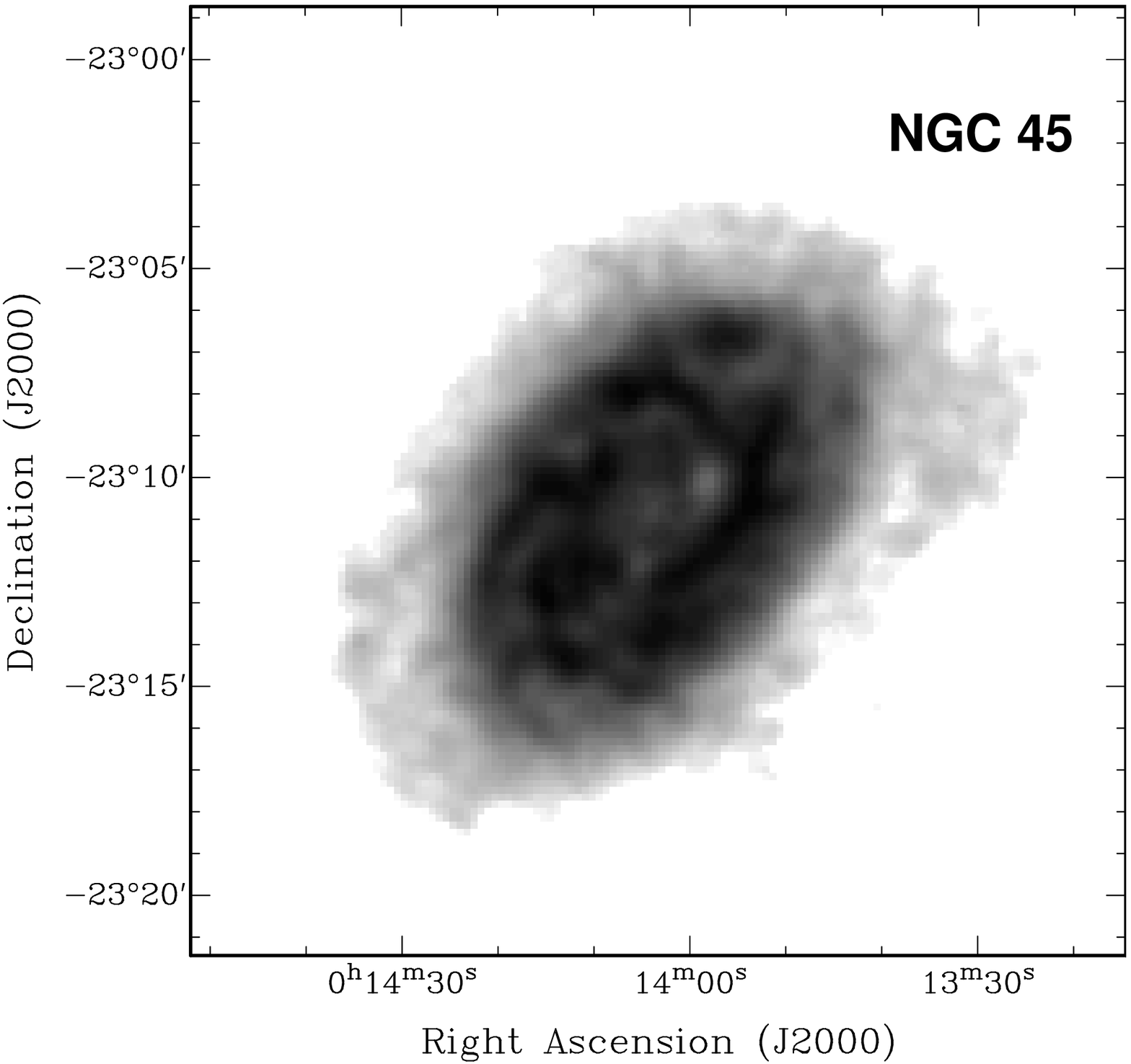}
\caption{NGC 45 : Contours of the DSS optical image (left). Greyscale total \hi map displayed using a logarithmic stretch (right).}
\label{contoptgreyhi_n45}
\end{figure*}

First, the {\it uv} datasets were carefully examined to
detect and reject bad points due to interference or
crosstalk between antennae. Antennae which were shadowed by
another antenna during the observations were flagged bad for the
shadowing period. The data were then calibrated using the
standard VLA calibration procedure. The flux scale was obtained
using the continuum source 0134+329. A bandpass calibration was also
applied. The reduction was done using the NRAO software package AIPS
at the VLA and at the Universit\'e de Montr\'eal. 

First, by creating
and inspecting a preliminary series of channel maps, the
channels containing only continuum radiation were 
identified. Next, those channels,
free of emission line, were averaged in the {\it uv} plane, 
to represent the continuum radiation, and
subtracted from all the channels. 
Finally, the channel maps were
produced and cleaned simultaneously  via a Fourier
transform with natural weighting and no taper. The pixel size was 10$''$ for
both systems and the original beam were convolved by circular beams of 40$''\times 40''$ for NGC 24 and
42$''\times 42''$ for NGC 45. This was done to make sure that all the structures that are observed are not due 
to the original slightly elongated beam shape.
The resulting maps had a rms noise of
2.4~mJy/beam and 1.6~mJy/beam for NGC 24 and NGC 45 respectively.
 
\section{Optical properties}
\label{sec:photometry}
\subsection{NGC~24}

 A study of  the $I-$band isophotes shows that the
intrinsic axis ratio varies from 0.5 near the center to 0.26 at the
last measured isophote.   The mean photometric parameters  
computed within the radius range $1.5'-3.0'$ are 
$i = 78 \pm 5\degr$ (b/a = 0.26) and $PA = 225 \pm 5\degr$. 

The elliptically averaged luminosity profile in the $I-$band, 
illustrated in Figure~\ref{lp} and listed 
in Table~\ref{Iprof-n24}, was obtained using 
the \textit{ellipse} task in IRAF\footnote{IRAF
is distributed by NOAO, which is operated by AU, 
Inc. under cooperative agreement with the NSF.}.
In the very inner parts ($R \le 50''$) the
profile shows a decrease, probably due to internal absorption, while in
the outer parts, it reveals a typical exponential decline.
An exponential fit to the $I-$luminosity profile (80$'' \le R
\le 180''$) yields an extrapolated central surface brightness of
19.12~mag\,arcsec$^{-2}$ which, when corrected for $I-$Galactic
extinction and inclination, gives $I(0)_c = 20.67$ mag arcsec$^{-2}$.
A galactic extinction $A_I = A_B/2.5 = 0.024$ (Draine 1989) was used, 
where $A_B=0.06$ according to de Vaucouleurs et al (1991) (hereafter
referred to as RC3). The derived exponential scale length is
$\alpha^{-1} \sim$ 43\arcsec, which corresponds to $\sim$ 1.42 kpc at 6.8 Mpc. 

When transformed to the $B-$band ($\mu _B = \mu _I + 1.452$ 
for  the Sc morphological type, from Carignan 1983\footnote{This was derived using mean Spectral Energy Distributions (Coleman et al. 1980) 
  for the different morphological types, convolved with the $UBVRI$ filter responses (Bessell 1979).}), this finally gives 
$B_c(0) = 22.12 \pm 0.3~{\rm mag\,arcsec^{-2}}$,  
which is faint relative to the Freeman canonical value of $21.65 \pm 0.30$  
for normal spiral galaxies (Freeman 1970). 
The central surface brightness of NGC 24
lies in the range of 22.07 $<$ $B_c(0)$ $<$ 23.70 mag, as defined by
Romanishin, Strom \& Strom (1983) for their LSB sample, but is at the limit (within the errors) between low surface brightness objects and normal galaxies.
One notices here that contrary to NGC 45, which is a genuine LSB (Romanishin et al. 1983, see \S\ref{photn45})  
with anemic spiral arms and no evident bulge (Fig.~\ref{dssimages}), 
the morphological properties of NGC 24 (presence of distinct spiral arms and a small bulge) are comparable with 
those of high surface brightness late-type spirals.  
 
Integrating
the $B-$luminosity profile gives a total
apparent magnitude
$B_T$ = 12.13, which is in good agreement with the value of 
12.10 given by Sandage \& Tammann (1987) and the value of 12.19 in RC3. 
When corrected for internal absorption
($A_i/2.5=0.38$ (Draine 1989) where $A_i=0.95$\, from RC3), it gives
 $B_T^{0,i}$ = 11.75. At the distance of 6.8 Mpc, the corrected
absolute magnitude is $M_T^{0,i}(B)$ = $-$17.41. The
corrected absolute magnitude corresponds
to a total blue luminosity of $1.4 \times 10^9~L_{B \odot}$. Table~\ref{opar-n24}
summarizes the optical parameters of NGC 24.


\subsection{NGC~45}
\label{photn45} 
The photometric inclination is $55\pm 5\degr$ 
and the position angle of the major axis is $145\pm 5\degr$
(Puche \& Carignan 1988). 
 
 A $B-$band surface brightness profile is used for NGC 45, as derived in Romanishin et al. (1983). 
 We refer the reader to this article for the observational information on these  data. 
The $B-$band profile is shown in Figure~\ref{lp} and listed in Table~\ref{Bprof-n45}. 
Romanishin et al. (1983) give an extrapolated 
central surface brightness $B_c(0) =  22.51 \pm 0.46$ mag\,arcsec$^{-2}$ and 
a scale length of $\sim$ 77\arcsec. This value translates into 2.20~kpc 
at our adopted distance of 5.9 Mpc.
Integration of the $B$-band profile gives a total apparent magnitude
of $B_T = 11.48$, which is comparable with the value of 11.10 given by Sandage \& Tammann
(1987) or the RC3 value of 11.32. 
This corresponds to an
absolute magnitude $M_T^{0,i}(B)$ = $-$17.45 and a total blue luminosity
of $1.5\times 10^9~L_{B \odot}$, at the distance of
5.9~Mpc. The optical parameters are summarized in Table~\ref{opar-n45}. 


\begin{figure*}\centering
\includegraphics[width=\columnwidth]{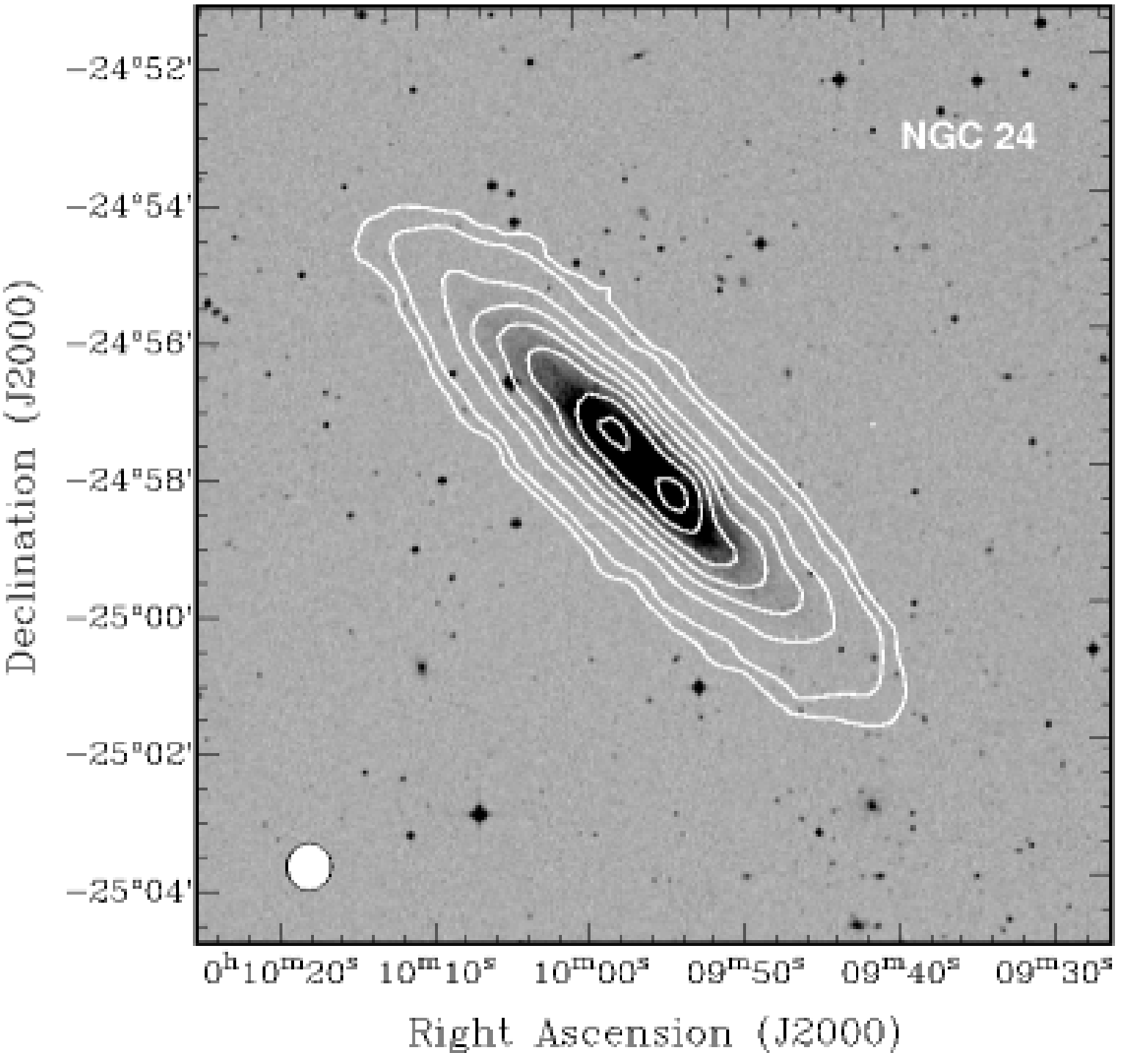}\includegraphics[width=\columnwidth]{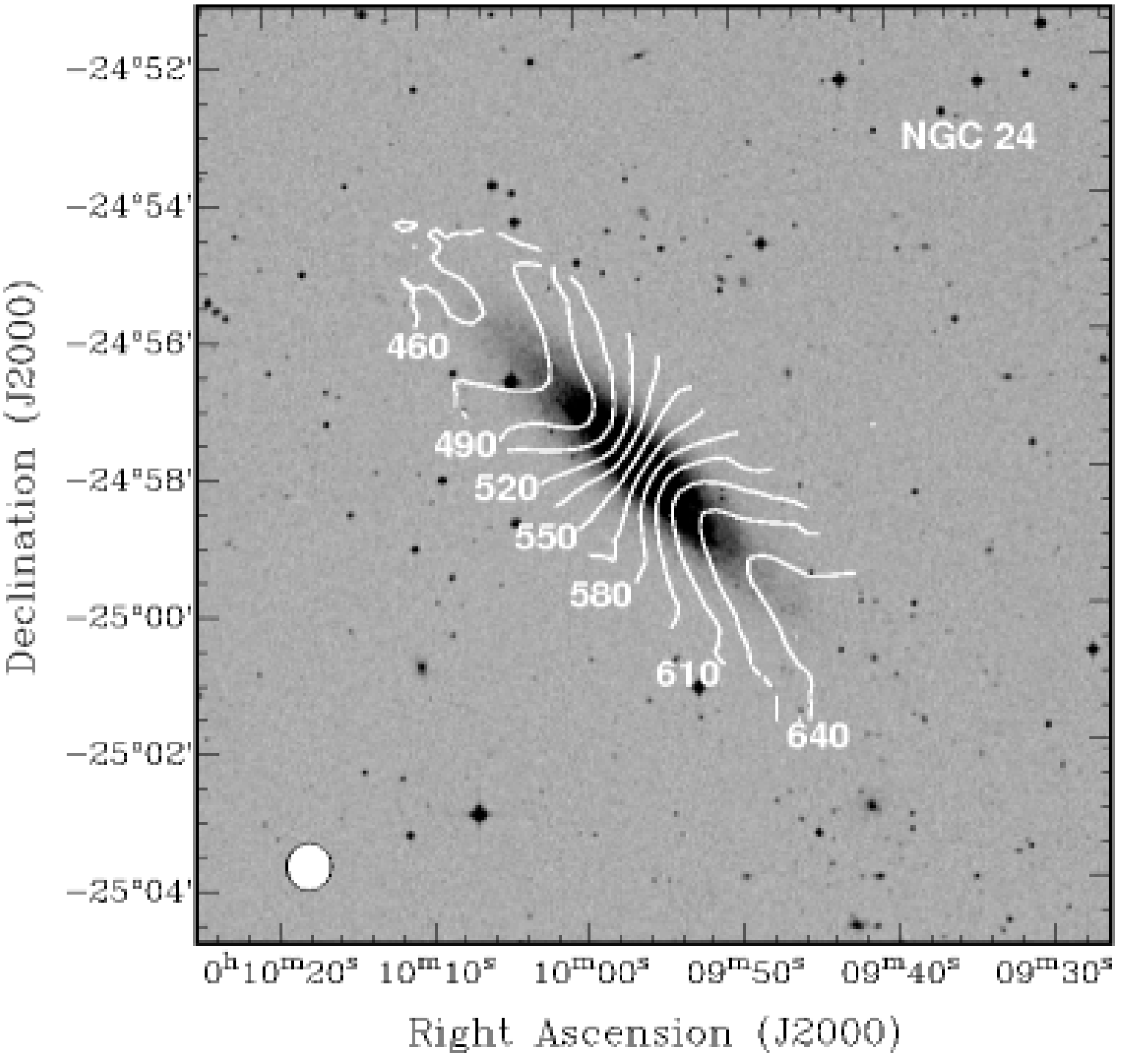}
\caption{NGC 24 : Total \hi map (left) and \hi velocity field (right) 
superimposed on a DSS image of the galaxy. The contours are at surface
density levels of 0.06, 0.31, 1.25, 2.19, 3.13, 4.07, 5.01 and $5.95\times 10^{21}{\rm~atoms\,cm^{-2}}$. 
Velocity contours are drawn from 460 to 640 \kms\ in steps of 15 \kms. The beam size of $40\arcsec \times 40\arcsec$
is shown in the bottom left corner of the images.
\label{momn24}}
\end{figure*}


\begin{figure*}\centering
\includegraphics[width=\columnwidth]{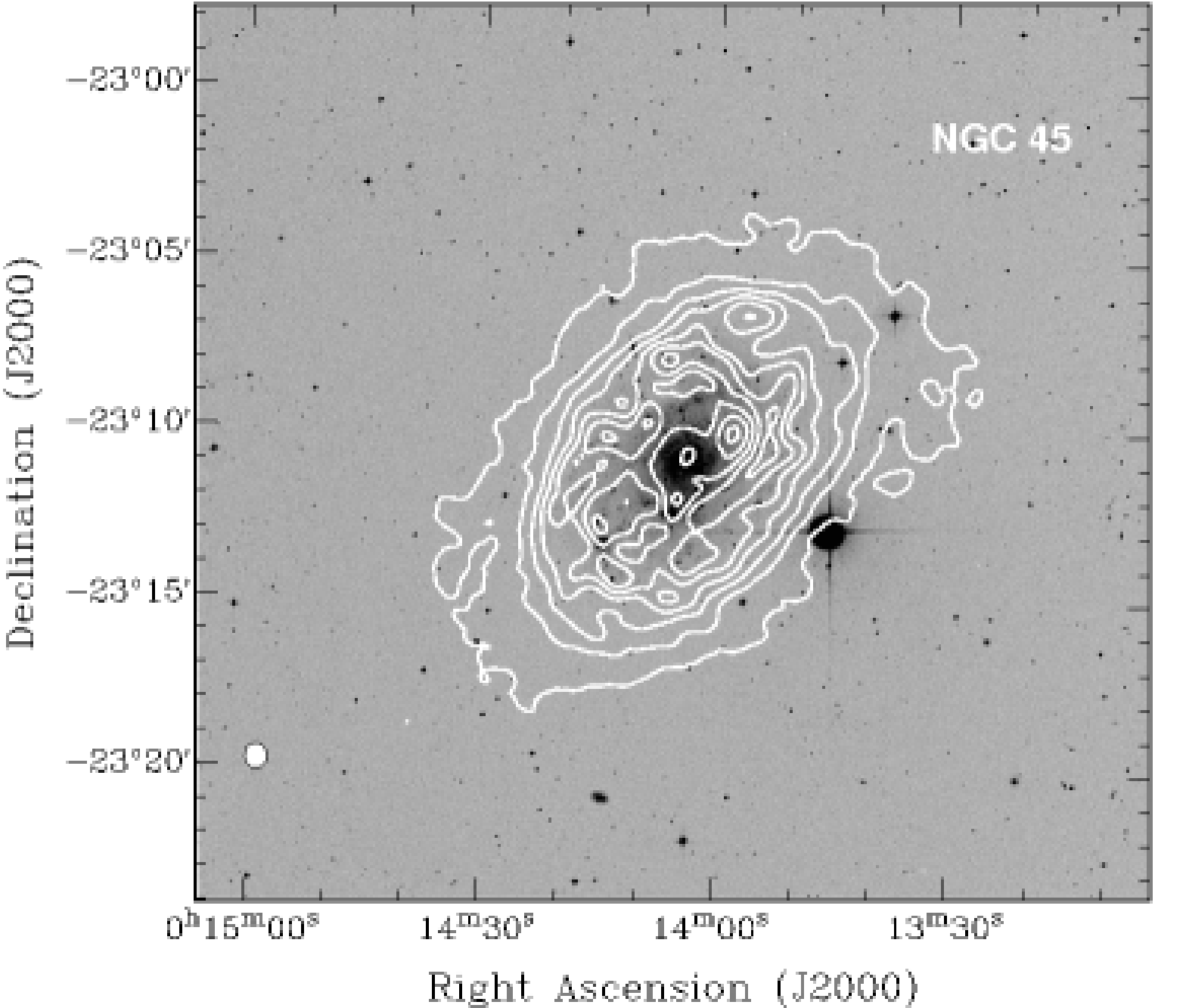}\includegraphics[width=\columnwidth]{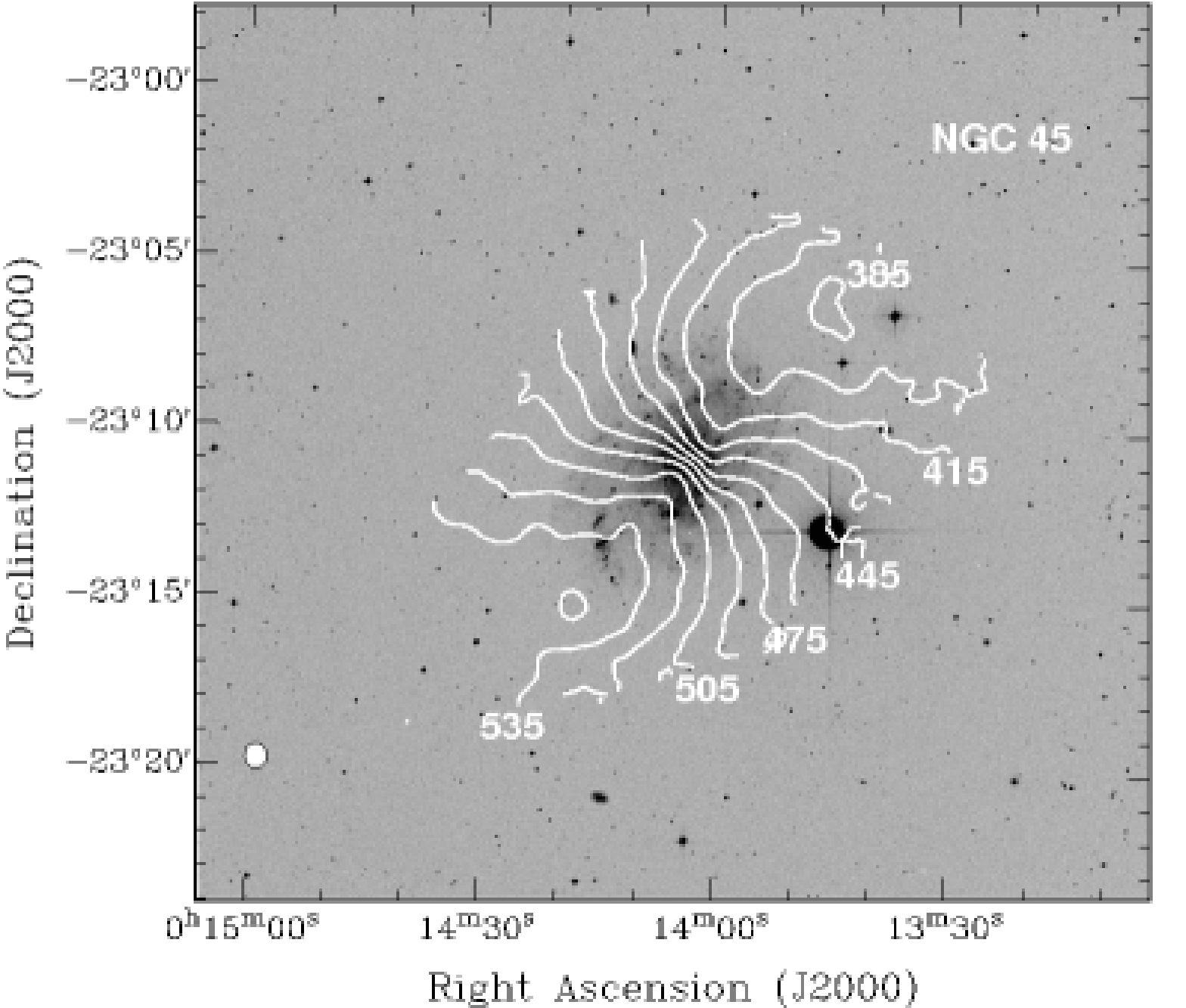}
\caption{NGC 45 : Total \hi map (left) and \hi velocity field (right) 
superimposed on a DSS image of the galaxy. The contours are at   surface
density levels of  0.13, 0.62, 1.11, 1.60, 2.09, 2.58 and $3.07\times 10^{21}{\rm~atoms\,cm^{-2}}$.
Velocity contours are drawn from 385 to 550 \kms\ in steps of 15 \kms. The beam size of $42\arcsec \times 42\arcsec$ is 
shown in the bottom left corner of the images.
\label{momn45}}
\end{figure*}

\section{\hi content and distribution}
\label{sec:neutrgas}
\subsection{NGC~24}

\subsubsection{Global Properties}

Once the correction for the primary beam attenuation had been applied,
the flux in each individual channel was summed to give the global \hi 
profile of  Figure~\ref{global} (left panel). An intensity-weighted systemic velocity of ${\rm
547 \pm 3}$ \kms and a midpoint heliocentric radial velocity of
$V_\odot = 549 \pm 3$ \kms were derived. 
The measured profile widths at 20\% and 50\% levels are $W_{20} =  
218 $ \kms and $W_{50} = 208$ \kms.
This can be compared with the HIPASS (Koribalski et al. 2004) values of
$V_\odot = 554$ \kms, $W_{20} = 223$ \kms 
and $W_{50} = 210$ \kms.


The integrated flux of $54 \pm 5$ Jy \kms\ is comparable with the 
HIPASS measurement of 50.3 Jy \kms (Koribalski et al. 2004).
It implies a total \hi mass of $(5.87 \pm 0.5) \times 10^8$ \msol, somewhat 
higher than the value of $(4.98 \pm 0.4) \times 10^8$ \msol
given by Huchtmeier \& Seiradakis (1985) for a distance of 6.8 Mpc. 
It appears that no flux was missed by our synthesis observations. 

\subsubsection{Spatial Distribution}
\label{sec:neutrgasn24}
\begin{figure*}\centering
\includegraphics[width=\columnwidth]{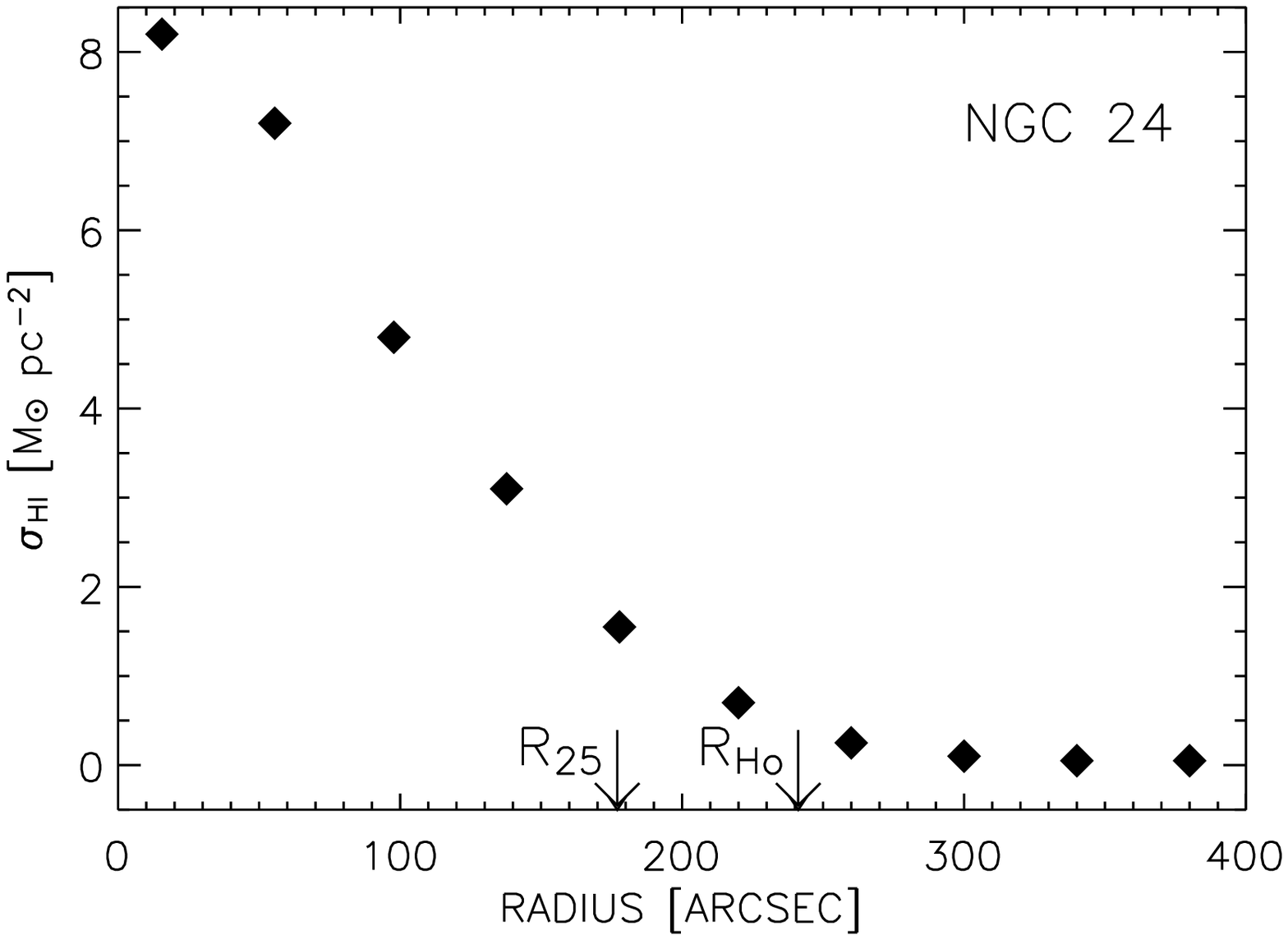}\includegraphics[width=\columnwidth]{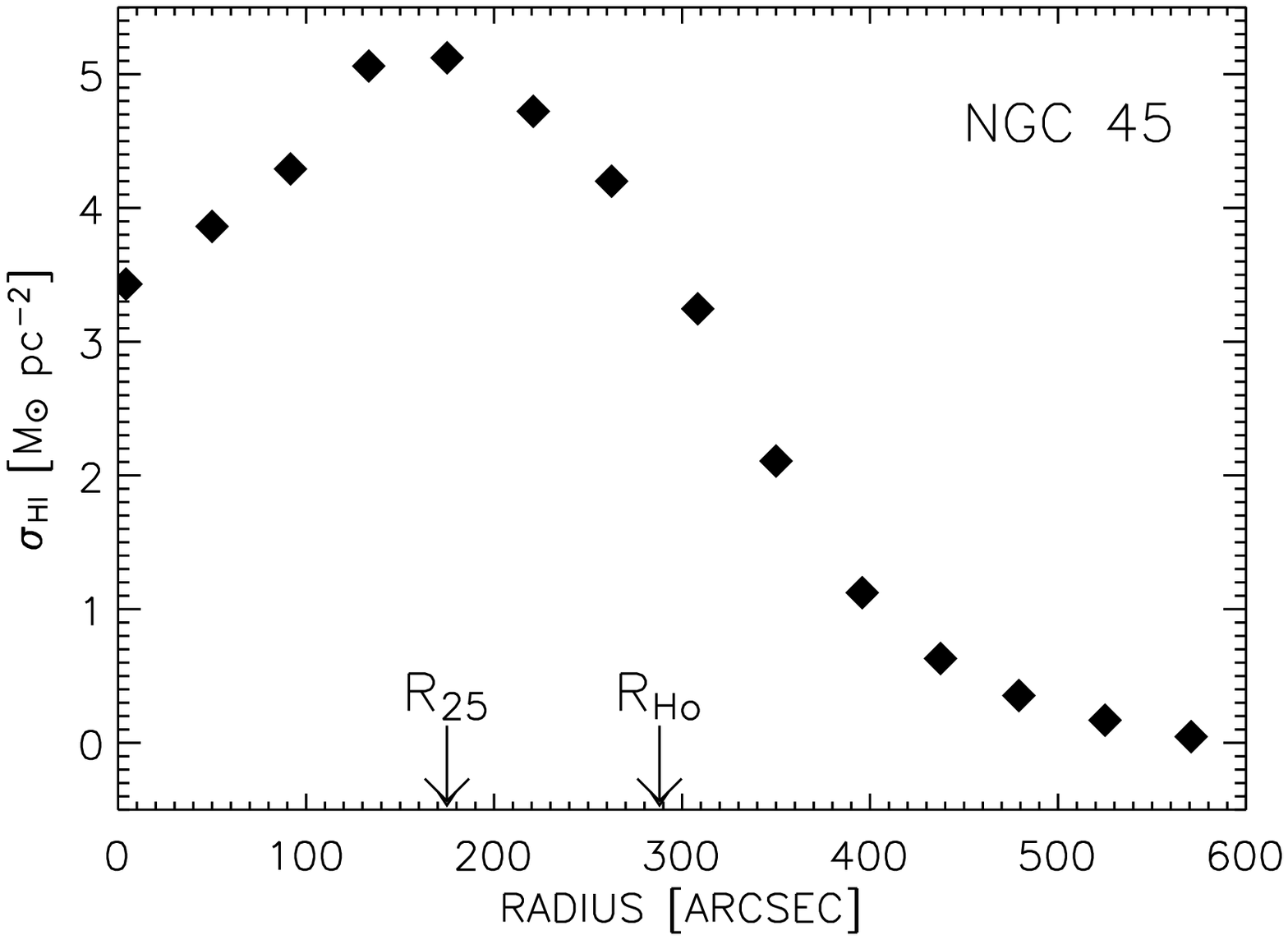}
\caption{Radial distribution of \hi surface density for NGC 24 (left)
and NGC 45 (right),
obtained by averaging the total \hi map in circular annuli in the plane
of the galaxy. 
\label{HIprof}}
\end{figure*}

A moment analysis (with the \textit{momnt} task in AIPS) 
produced the total \hi emission map
of Figs.~\ref{contoptgreyhi_n24} (right panel, in greyscale) and~\ref{momn24} 
(left panel, expressed as column densities contours and superimposed
on a DSS image of the galaxy). The distribution, which is
really symetrically distributed, stretches out to $\sim$ 10\arcmin\ in diameter 
$(\sim 1.3~{\rm R_{Ho}}$ at a level of $\sim 10^{20}~{\rm atoms\,cm^{-2}})$.
Concentric elliptical averaging corrected by
a factor $\cos{i}$ gave the \hi radial profile illustrated in 
Figure~\ref{HIprof} (left panel). 

The \hi surface density peaks at the center of the galaxy and then decreases as a function of radius.
As compared with other morphological types of galaxies (see Fig.~10 in Swaters et al. 2002), 
 this profile is more typical of Sd spirals (in shape and amplitude) than LSBs.  
 Both the morphological, optical and \hi properties of NGC 24 point out that this spiral can 
 be considered at the transition between normal and LSB galaxies.
This distribution will be used in Sec.~\ref{sec:mass} to derive the dynamical
contribution of the \hi disk. 


\subsection{NGC~45}

\subsubsection{Global Properties}

The \hi properties of NGC~45 were derived in a similar way as  
for NGC~24. Figure~\ref{global} (right panel) gives the global \hi profile once the correction
for the primary beam attenuation was applied. The total flux density
of $186 \pm 19$ Jy \kms 
(to be compared with $195.8$ Jy \kms in HIPASS; Koribalski et al. 2004)
corresponds to an \hi mass of $(1.52 \pm 0.2) \times 10^9$ \msol, 
which is to be compared with $(1.98 \pm 0.1) \times 10^9$ \msol 
(Huchtmeier \& Seiradakis 1985) for a distance of 5.9 Mpc. 
This seems to indicate that $\sim$
20\% of the flux is missing, probably due to missing
short spacings in the VLA observations.
Also derived from the profile is the midpoint systemic velocity of
$470 \pm 3$ \kms and the intensity-weighted
systemic velocity of $473 \pm 3$ \kms. 
The measured profile widths are $W_{20} = 180$ \kms and $W_{50} = 167$ \kms. 
This can be compared to the HIPASS values of
$V_\odot = 467$ \kms, $W_{20} = 185$ \kms
and $W_{50} = 167$ \kms (Koribalski et al. 2004).

\subsubsection{Spatial Distribution}

Figures~\ref{contoptgreyhi_n45} (right panel, in greyscale) and~\ref{momn45} (left panel, in contours superimposed
on a DSS photograph of the galaxy) show the distribution of \hi surface densities obtained by the moment analysis.
 The \hi disk is regular and extends to $\sim$ 20\arcmin\ in diameter $(\sim 2.1~{\rm R_{Ho}}$ 
at a level of
$\sim 10^{20}~{\rm atoms\,cm^{-2}})$. 

The shape and amplitude of the radial distribution of the \hi 
surface density (Fig.~\ref{HIprof}, right panel) is 
typical of what is seen in low surface brightness spirals, 
showing a depression in the central regions with a low gas surface density level 
(see Fig. 10 in Swaters et al. 2002 for comparison). 
Both morphological, photometric and gaseous properties of 
NGC 45 make this  spiral a genuine low surface brightness galaxy.
This profile will be used in Sec.~\ref{sec:mass} to
evaluate the dynamical importance of the NGC 45 \hi component.

\section{Velocity field and rotation curve}
\label{sec:kinematics}
\subsection{NGC~24}

Figure~\ref{momn24} (right panel) shows the velocity field obtained by the moment
analysis where the radial velocities were calculated by taking the
intensity weighted mean of each line profile for pixels above the 1.8$\sigma$
level, in the 40\arcsec$\times$40\arcsec\ resolution data cube. This velocity
field is regular, with no sign of large-scale deviation from axial
symmetry. From the shape of the isovelocity contours, one can 
already infer that this galaxy does not have a solid-body rotation
curve (the contours are not parallel) but rather tends  
to be flat in the outer parts.


\begin{figure*}\centering
\includegraphics[width=5.3cm]{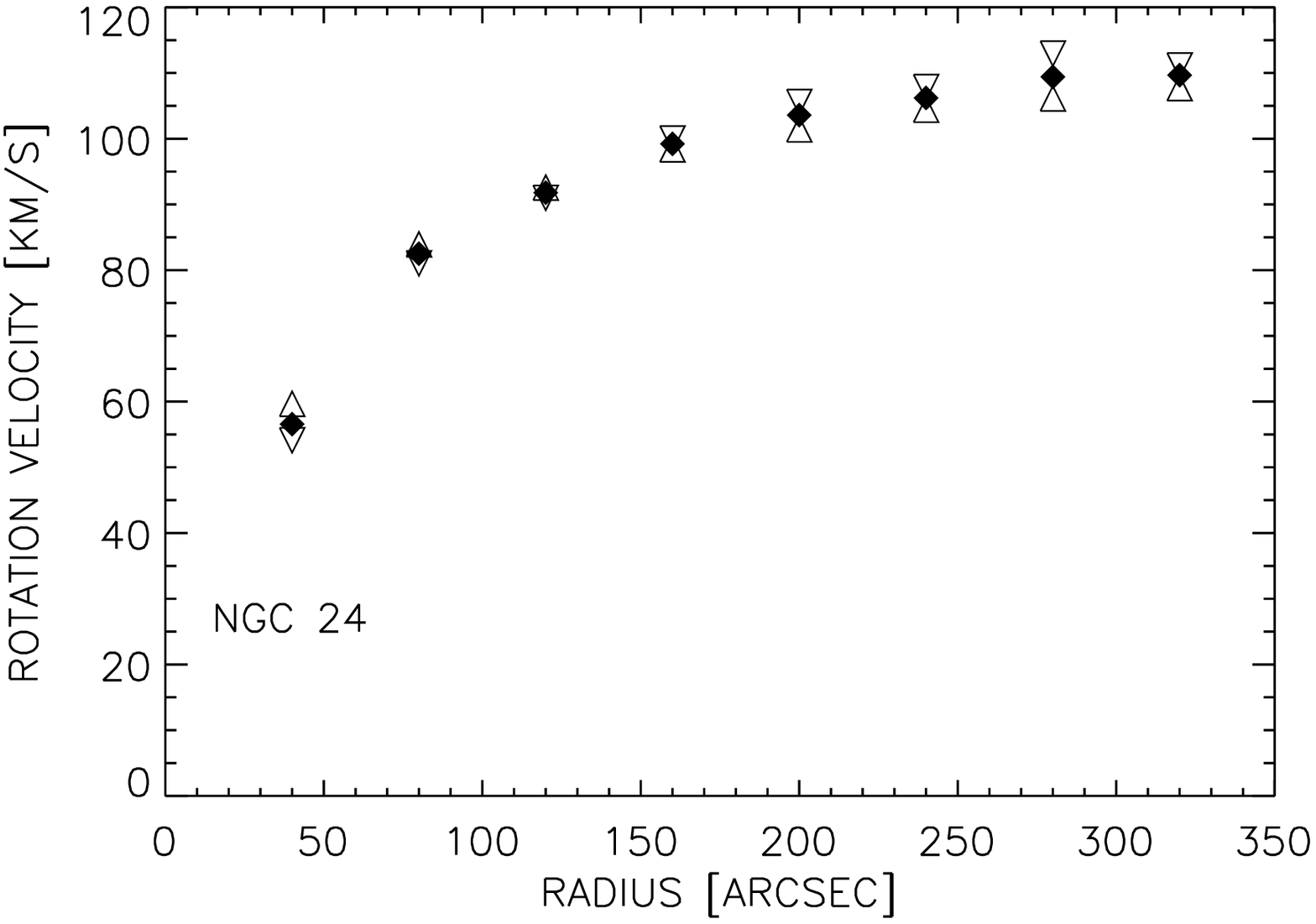}\includegraphics[width=5.3cm]{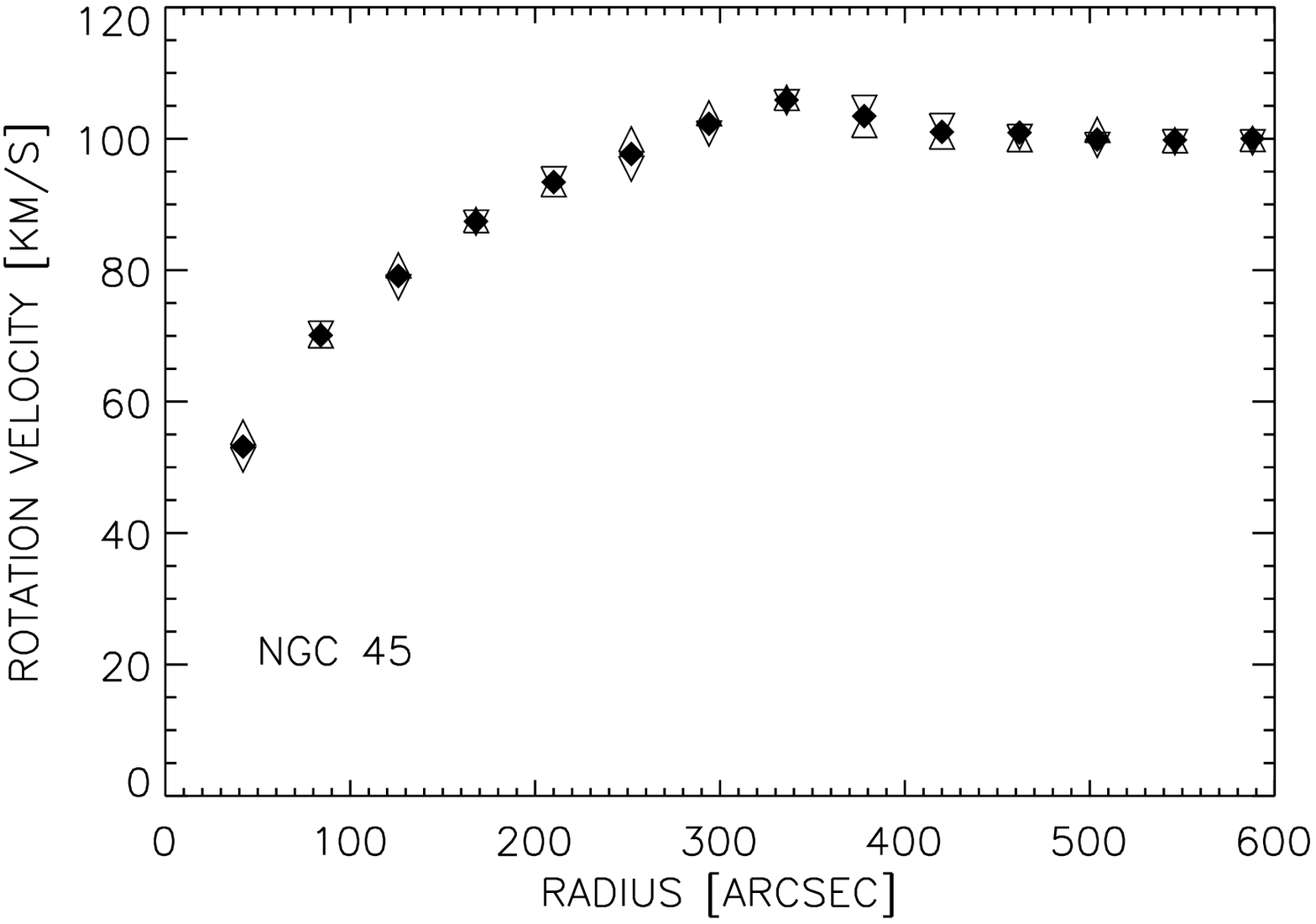}\includegraphics[width=5.3cm]{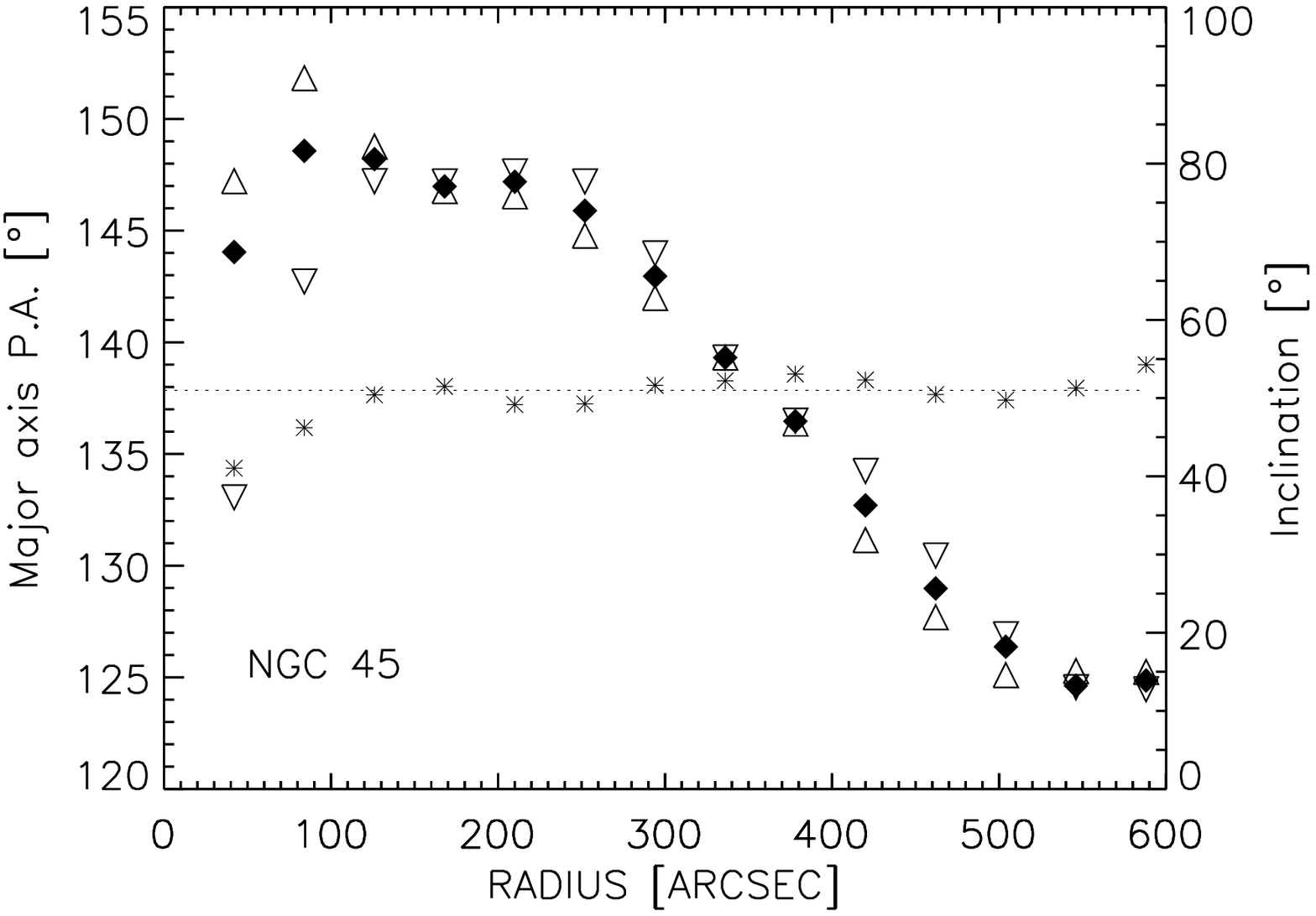}
\caption{Rotation curves of NGC 24 and NGC 45 (left and central panels resp.) 
and radial profile of the inclination and the major axis position angle
of NGC 45 (right panel), as derived from the tilted-ring models. 
Filled diamonds correspond to the values for both sides fitted 
simultaneously, open upward and downward triangles 
to the values for the receding and approaching sides 
(respectively) fitted separately. The inclination is shown 
as star symbols and the adopted one as a dashed line. \label{rc}}
\end{figure*}

In order to extract the rotation curve (RC), the \textit{rotcur} task (Begeman 1989) of the GIPSY program (van der Hulst et al. 1992) was 
used. The task makes a least-square fitting of a tilted-ring 
model to the velocity field. A solution has to be found for the following five
kinematical parameters: the coordinates of the rotation center, the systemic velocity ($v_{\rm sys}$), the
inclination  ($i$) and the position angle ($P.A.$) of 
the major axis in order to obtain the
circular velocity as a function of radius.
To diminish the importance of deprojection errors at such an
inclination, radial velocities in an opening angle
of 80$\degr$ about the minor axis were rejected for the least-square
fitting and a cosine weighting function giving more
importance to points near the major axis was used.

The fitting procedure is described in Chemin et al. (2006). 
Briefly, the rotation center and systemic velocity are first searched for, then the 
inclination and the position angle.  The derived 
systemic velocity is $554 \pm 1$ \kms, which is comparable
to the value obtained with the global \hi profile.
The kinematical inclination is $64\pm 3\degr$. 
This is significantly lower than the optical inclination of $78^\circ$ and  
the  value found by an ellipse fitting to the \hi isophotes  ($76\degr$). 
The low spatial resolution of the data  combined with 
  the high inclination  of the disk probably explains such a difference. 
The choice of either the photometric or the kinematical value does not influence the main result 
of this work (see Sections ~\ref{sec:mass} and~\ref{sec:discussion}). 
Hence, the kinematical inclination is chosen for NGC 24 throughout the article and all derived quantities 
(rotation curve, mass-to-light ratio, ...) are given adopting this value, except where explicitely mentionned.
The kinematical position angle is $229\pm 1\degr$. 
This value remains comparable with the one derived from the photometry ($225 \pm 5\degr$).

A rotation curve is finally obtained (Figure~\ref{rc} and Table~\ref{rc-n24}) 
by fixing all the other parameters  at constant values 
because no warp is found in NGC 24.
At each radius, the quoted error-bar of the velocity point given 
in Table~\ref{rc-n24} corresponds to the maximum value 
between the formal error calculated by \textit{rotcur} 
($1\sigma$ dispersion of the fitted velocity parameter) 
and the largest velocity difference between the solution for both sides
and the separate solutions for the approaching and receding halves (Carignan \& Puche 1990a).  

\subsection{NGC~45}

The velocity field of NGC 45, obtained from the analysis of the 
42\arcsec$\times$42\arcsec\ resolution data cube by discarding pixels under a
1.6$\sigma$ level, is shown in Figure~\ref{momn45} (right panel).  

The same procedure as for NGC 24 is used to derive the RC of NGC 45. 
To lessen the errors due to the deprojection, all points in a sector
of 60$\degr$ around the minor axis were discarded from the fitting.  
A systemic velocity of 467$\pm$2\kms is found, which agrees
within the errors with the value obtained using the global profile  and
with the one given by Adler \& Liszt (1989). The inclination is 
found to be constant as a function of radius ($51 \pm 2\degr$), and
is in relative agreement with the photometric value within the error ($55\pm 5\degr$).

A study  of the \hi  isophotes shows that the outermost contour is 
slightly twisted  with respect to the inner isophotes. 
The outer isovelocity contours 
are also twisted as a function of radius. 
A kinematical twist of the \hi plane is indeed 
detected with the tilted-ring model (Figure~\ref{rc}). The $P.A.$ 
decreases by $\sim 25\degr$ from the inner to the outer regions 
whereas $i$ remains constant. 
A kinematical warp  or a simple twist  of the kinematical major axis 
is very common in the Sculptor group 
\hi disks (e.g. Carignan \& Puche 1990a), 
and more generally in spiral galaxies (Garc\'ia-Ruiz, Sancisi \& Kuijken 2002).  

The final rotation curve is thus derived  by fixing the center coordinates, systemic velocity and inclination 
at constant values and by leaving the position angle free as a function of radius. 
The RC is given in Table~\ref{rc-n45} and displayed in Figure~\ref{rc}. 
Here again, no significant asymmetry is detected between the RCs of 
the approaching and receding sides of the disk.  

Table~\ref{rc-n45} also gives the radial variation of the $P.A.$ with its errors computed as 
 the maximum value between the formal error calculated by \textit{rotcur} 
($1\sigma$ dispersion of the fitted angle) and the largest angle difference between the fitted value for both sides
and the separate fitted values for the approaching and receding halves.

\section{Study of the mass distribution}
\label{sec:mass}

A preliminary study of the mass distribution is 
presented here using the present 
low resolution \hi data. This should
give a first good estimate of the dark-to-luminous mass ratio in
those two galaxies and allow us to get a good idea 
whether the dark matter component
dominates at all radii, as seen in dwarf irregulars 
(e.g. Carignan \& Beaulieu 1989), 
in some late-type spirals (e.g.  C\^ot\'e et al. 1991) or 
in other LSB galaxies (e.g. de Blok \& McGaugh  1997).
As for the other papers from this series,
only the best-fit mass models are presented for the two galaxies. 
These models are very close to the maximum disk case.   
Notice that for NGC 45 another model that uses the 
mass-to-light ratio expected from 
stellar populations synthesis (SPS) models 
is presented (see Sec.~\ref{sec:discussion}).

 
\begin{figure*}\centering
\includegraphics[width=5.3cm]{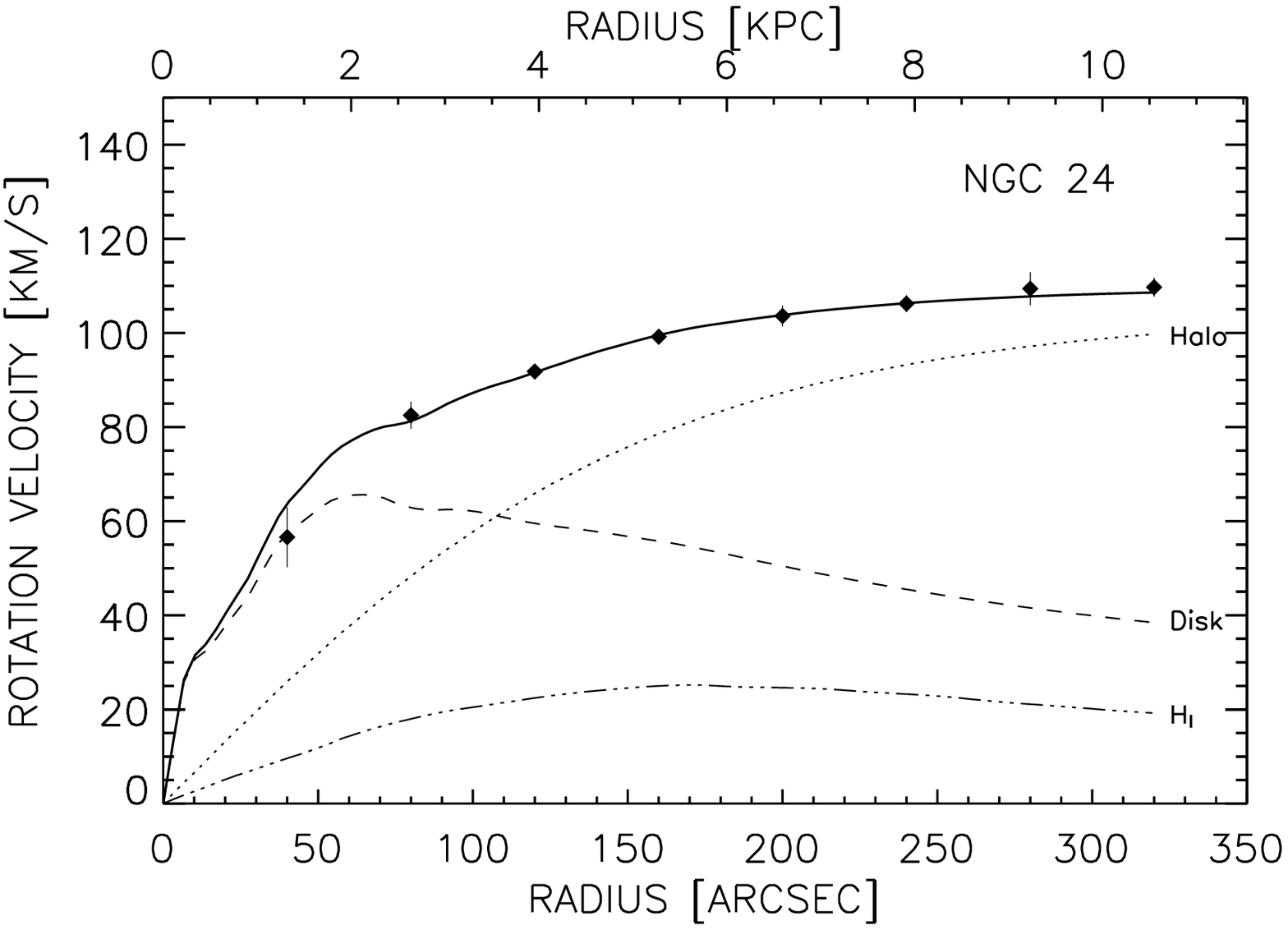}
\includegraphics[width=5.3cm]{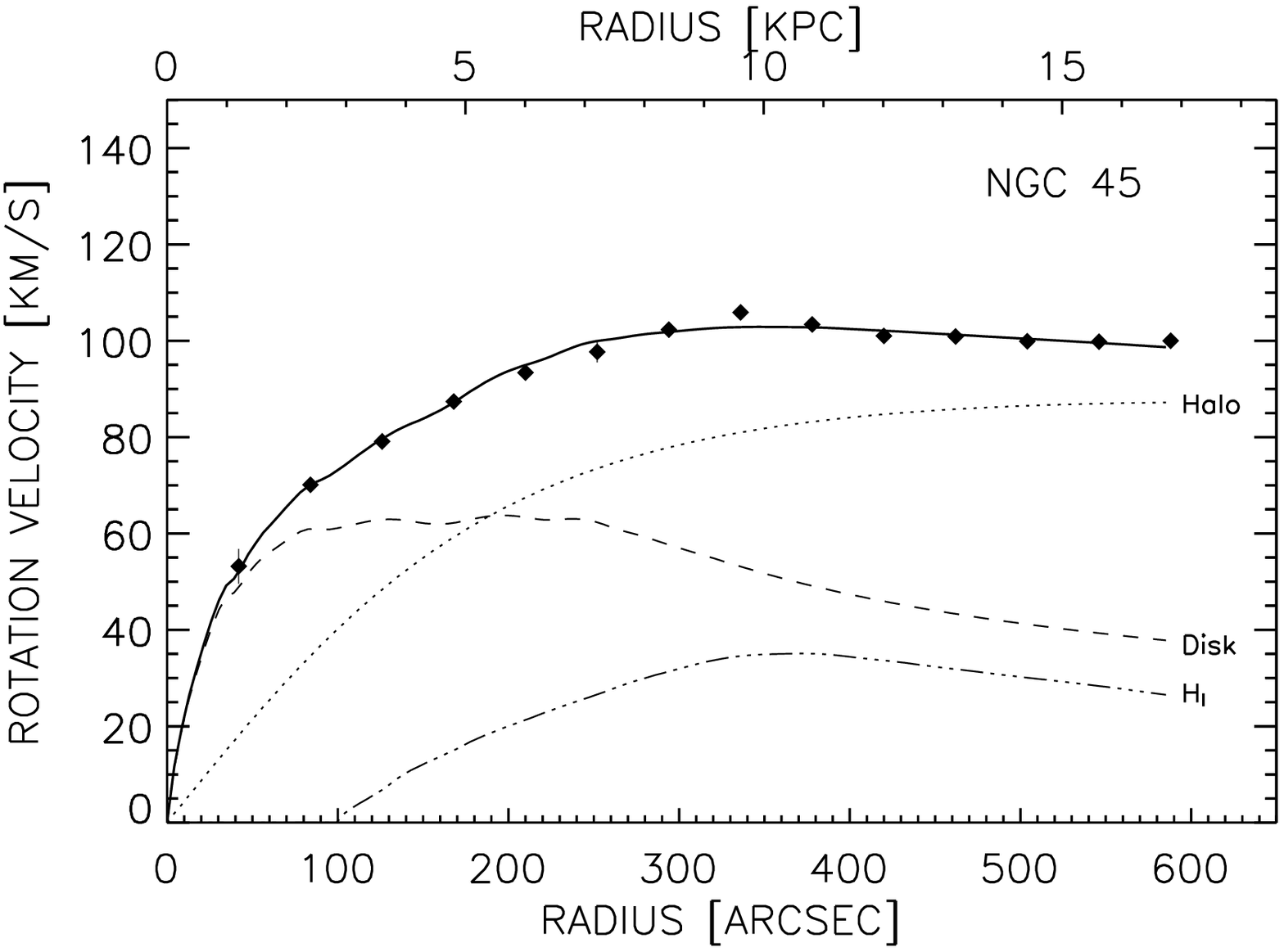}
\includegraphics[width=5.3cm]{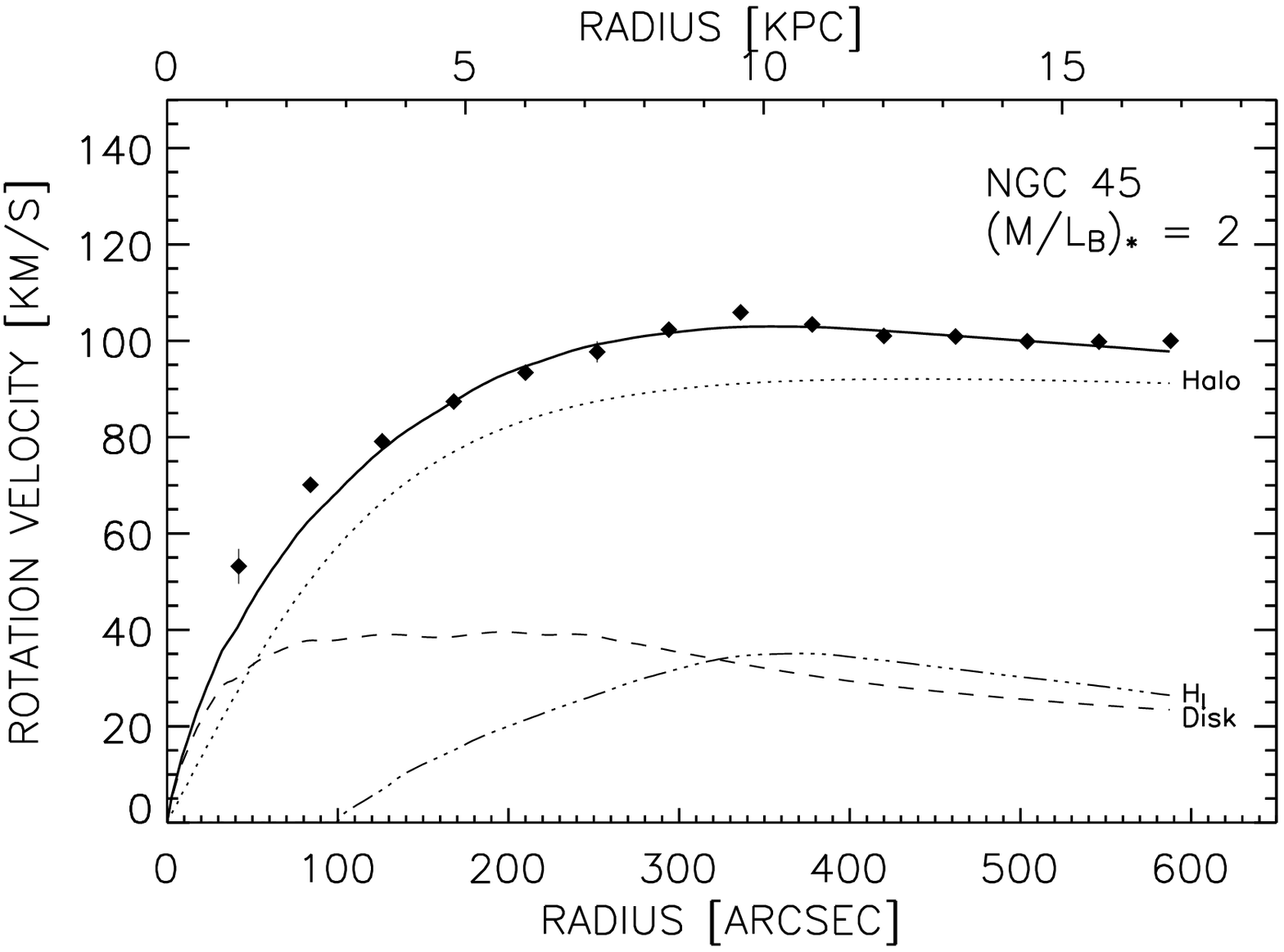}
\caption{Mass models using the best-fit method for NGC 24  
and NGC 45 (left and central panels resp.). The model on the right panel is for NGC 45
with \rml = 2.0. The full line is for the 
best-fit model to the data (filled diamonds). 
The contribution of the different components (stellar disk,
\hi component corrected for the He content \& dark halo)
are identified. \label{mm}}
\end{figure*}

\subsection{Two Component Model}

One can refer to Carignan (1985), Carignan \& Freeman (1985) 
and the other papers in this series for a detailed discussion of the two
components (dark and luminous) model.
Because NGC 24 has a very small
bulge and NGC 45 has no bulge, as seen in Fig.~\ref{dssimages}, 
no attempt to include a bulge component in the models was made. 
For NGC 24, the $I-$profile derived in Sec.~\ref{sec:photometry} was 
used to derive the contribution of the stellar component. 
It was transformed into the $B-$band according to Carignan (1983). 
For NGC 45, the B-band luminosity profile of Romanishin et al. (1983) was used.
The contribution of the gaseous component was derived 
assuming that all the gas is confined 
in an infinitely thin disk  and using the \hi radial 
surface densities (Fig.~\ref{HIprof}), which 
were multiplied by 4/3 to take into account primordial helium.

The dark halo is modeled by an isothermal sphere
which is described by two basic parameters: the core
radius $r_c$ and the one-dimensionnal velocity dispersion $\sigma$. A
third quantity, the central density of the halo, is related to the two
others by $\rho_0 = 9\sigma^2/{4\pi Gr_c^2}$. Essentially, the mass
model depends on three parameters: the amplitude scaling of the
luminous disk \rml (the mass-to-light ratio of the stellar disk), 
the radial scaling $r_c$ and the velocity scaling $\sigma$ of the halo.
To determine the combination of the three parameters which best reproduces
the observed rotation curve, a best-fitting method was used, without
any constraints on the parameter values. By
exploring a grid of values in the three parameter space, a set
[\rml, $r_c$, $\sigma$)] is found leading to the smallest $\chi^2$ for the
fit. Once an approximate minimum has been identified, the solution is
refined by improving the step resolution for the three parameters. This routine 
is reiterated until a final set of parameters is obtained. 
The mass-to-light ratio of the stellar disk is supposed to be constant as a function of radius  
and the errors on the derived model parameters are established from the 90\%
confidence level for both galaxies.

Notice that no attempt to explore different functional forms for the halo was made with these \hi data. 
A comparison between a cuspy halo, like e.g. the NFW one (Navarro, Frenk \& White 1997), 
 and a core-dominated halo, like the one presented here, indeed requires 
 high spatial resolution data for accurately mapping the inner rising part of a rotation curve 
 (Swaters, Madore \& Trewhella 2000). Such mass models will be presented elsewhere when 
 three-dimensional optical Fabry-Perot data  will become available (Chemin et al., in prep.). 
 
\subsection{Mass modeling results}

The results from the best-fit mass model for NGC 24 are illustrated in 
Figure~\ref{mm} (left panel) and given in
Table~\ref{mod-n24}. The  model gives a \rml of 2.5 $\pm$ 0.3 for a
total disk mass of $\sim 3.4 \times 10^9$ \msol. 
The \hi + He gas, with a total mass
of $\sim 7.4 \times 10^8$ \msol, 
provides only $\sim$20\,\% of the luminous mass and
is therefore not very important dynamically. The parameters for the dark halo
are $r_c = 5.6 \pm 1.5$ kpc and
$\sigma = 65.0\pm 4.0$ \kms. This gives a central density for the
dark halo of  0.022 \msol${\rm pc^{-3}}$. 
 Notice here that the use of a rotation curve derived by a tilted-ring model with  
 an inclination fixed to the photometric value of 78\degr\ would slightly reduce the 
    \rml to 2.0 without changing the other parameters.  

The results from the best-fit mass model for NGC 45 are illustrated in
Figure~\ref{mm} (middle panel) and given in
Table~\ref{mod-n45}.
The parameters are \rml $= 5.2 \pm 1.0$, $r_c =
6.2 \pm 0.6$ kpc and $\sigma = 55.0 \pm 1.0$ \kms, 
which correspond to a central density for the
halo of $\rho = 0.013$ \msol$\rm pc^{-3}$. 
Total masses of $5.3 \times
10^9$ \msol and $2.1 \times 10^9$ \msol are found for the stellar    
and gaseous disks respectively.


\section{Discussion}
\label{sec:discussion}
 
The extent of the \hi disk  is ${\rm R_{HI}} = 1.3 \times {\rm R_{Ho}}  = 
7.3 \times \alpha^{-1}$ for NGC 24 
and  ${\rm R_{HI}}  = 2.1 \times {\rm R_{Ho}}  = 7.9 \times \alpha^{-1}$ for NGC 45. 
The \hi extent of NGC~24 is very similar to the \hi
extent of the Sculptor late-type spiral galaxies 
NGC 247, NGC 300 and NGC 7793 (1.2, 1.5 and 1.4$~{\rm R_{Ho}}$, respectively).
Again, this suggests that NGC 24 is very similar to normal
late-type spiral galaxies, as also claimed in \S\ref{sec:neutrgasn24}.
 
The best-fit mass model of NGC 24 gives a \rml 
for the stellar disk of 2.5 $\pm$ 0.3.
From the observed colors of 0.56 $\la (B-V) \la$ 0.60 for NGC 24 
(de Vaucouleurs et al. 1991), SPS models
predict a mass-to-light ratio for the stellar disk of 
1 $\la$  \rml $\la$ 1.2 (Bell \& de Jong 2001).  
The mass-to-light ratio is thus twice the one expected
from SPS models. This result remains unchanged when using 
\rml $= 2.0$, as found when using the photometric inclination instead of the kinematical value.

The case of NGC 45 is even more problematic since the derived \rml 
from the best-fit model is nearly 3 times greater than 
the value expected from SPS models: 5.2 versus 1.7 $\la$  \rml $\la$ 2.2 
(Bell \& de Jong 2001), based on the observed 
$(B-V)$ color of 0.71 $\pm$ 0.03 (de Vaucouleurs et al. 1991).

Similar high \rml values are often found in LSBs when fitting  maximum disk 
models (de Blok \& McGaugh 1997; de Blok, McGaugh \& Rubin 2001; de Blok \& Bosma 2002; Swaters et al. 2003). 
An interpretation of these results is that the maximum disk hypothesis 
must not hold for LSB galaxies (de Blok et al. 2001).  
As a consequence, these galaxies appear to be dominated by 
dark matter at all radii  when mass models preferentially use a mass-to-light ratio consistent with SPS models 
(de Blok \& McGaugh 1997; de Blok et al. 2001; de Blok \& Bosma 2002). 

Such a model is illustrated  for NGC 45 in Figure~\ref{mm} (right panel), 
using a \rml of 2.0. This value is chosen to 
be representative of the expected \rml range given by the SPS models (see above). 
Indeed, it can  be seen that the dark component is dominant over 
almost the whole stellar disk, 
though with the noticeable exception of the very innermost regions. 

This new model nevertheless highly underestimates the first two points of the RC, 
thus giving a worse fit ($\chi ^2 \sim 8$) than the best-fit model ($\chi ^2 \sim 3$). 
The quality of the fit also decreases as one goes towards the minimum 
disk hypothesis (\rml $=0$ / $\chi ^2 \sim 11$). 
Such a result is in agreement with what is found for other LSBs, 
as illustrated in de Blok \& Bosma (2002). 
For several galaxies from their sample, the minimum disk hypothesis, 
or a model using a low \rml, does not always provide a better fit than the maximum disk hypothesis. 

One finally notices that higher \rml values should be expected if a correction for beam-smearing 
was applied or if a higher angular resolution was used for the innermost velocity points 
of the curves. Indeed, it should give more steeply rising rotation curves in their inner parts than 
the ones presented here (see e.g. Swaters et al. 2000). 
This would worsen the discrepancy found between the low \rml values from SPS models 
and the maximum disk hypothesis. 
Therefore, if one finally admits that the maximum disk hypothesis is ruled out for LSBs 
(as favored in de Blok et al. 2001), 
 a more cuspy halo than the pseudo isothermal sphere should perhaps be used to 
better fit the inner velocity points of NGC 45 with a low \rml. 
This claim will be tested in a forthcoming article. 

\section{Summary and conclusions}
\label{sec:conclusion}
An optical and \hi study of NGC 24 and NGC 45 has been
presented. The main results are as follows:

(1) From the surface photometry of NGC 24 and NGC 45,
it is found that while NGC 24 is on the faint side for
normal galaxies, NGC 45 can be considered as a bona fide LSB galaxy. 
However, both galaxies have very similar absolute magnitudes $\sim -17.4$.

(2) The \hi distribution for NGC~24 and NGC~45 
extends to $\sim 1.3$ and 2.1$~{\rm R_{Ho}}$ (respectively) at a
level of $\sim 10^{20}~{\rm atoms\,cm^{-2}}$. 
The \hi extent of NGC~24 is very similar to the \hi
extent of the Sculptor late-type spiral galaxies 
NGC 247, NGC 300 and NGC 7793 (1.2, 1.5 and 1.4$~{\rm R_{Ho}}$, respectively). 

(3) The overall velocity fields of the two galaxies are very regular.  
NGC 45 exhibits a twist of the kinematical major axis, showing a $\sim$25\degr\ variation of its 
major axis position angle as a function of radius. 
However, its disk does not have a classical warp since the inclination 
remains constant. 

(4) The rotation curves derived from the velocity fields 
rise slowly and flatten at a velocity of $\sim 110~{\rm km\,s^{-1}}$ and
$\sim 100~{\rm km\,s^{-1}}$ respectively. It extends out to $\sim$11~kpc 
for NGC 24 and to $\sim$17~kpc for NGC 45, which corresponds to
$\sim 7.5$ scale lengths in both cases.

The rotation curves, combined with the luminosity profiles,
were used to study the mass distribution of NGC 24 and NGC 45.
Using a best-fit model, the main results are :

(1) \rml = 2.5, $r_c = 5.6~{\rm kpc}$ and $\sigma = 65~{\rm
km\,s^{-1}}$ for NGC 24, and \rml = 5.2, $r_c = 6.2~{\rm
kpc}$ and $\sigma =55~{\rm km\,s^{-1}}$ for NGC 45. 

(2) In both galaxies, the dark halo is the main contributor to the
total mass of the galaxies with a contribution of more than $\sim$80\,\% 
at the last measured point. Since the best-fit models are close to the
maximum disk case, this can be considered a lower limit.

(3) The \rml of 5.2 found for NGC 45 
and 2.5 for NGC 24 are high when compared with the values predicted by
 stellar population synthesis models for galactic disks of same colors as NGC 24 and NGC 45. 
This result is similar to what is seen in other LSB galaxies.
When a model for NGC 45 uses a \rml of 2.0, which value is adopted from SPS models, 
it allows the galaxy mass to be entirely dominated by the dark component,   
but it also severely underestimates the velocity of the innermost 
points of the rotation curve.  

This article is the first from a series that aims at measuring the shape 
of the mass density profile for the dark component of NGC 24 and NGC 45. 
When higher resolution optical kinematical \ha\ data 
obtained with Fabry-Perot interferometry become available,
more accurate mass models of those galaxies will be presented. 
 
\acknowledgments
We would like to thank the Very Large Array for allocations of telecope
time and the NRAO staff for valuable assistance. We are very grateful to Sylvie 
Beaulieu for her help. LC acknowledges partial support
from the Fonds Qu\'eb\'ecois de la Recherche sur la Nature et les Technologies
and CC from the Conseil de Recherches en Sciences Naturelles et en
G\'enie du Canada. N.D. ackowledges financial assistance from Fonds FCAR Qu\'ebec. 
The Digitized Sky Surveys were produced at the Space Telescope Science Institute under U.S. 
Government grant NAG W-2166. The images of these surveys are based on photographic data 
obtained using the Oschin Schmidt Telescope on Palomar Mountain and the UK Schmidt Telescope. 
The plates were processed into the present compressed digital form with the permission of 
these institutions.

\clearpage
 

\begin{deluxetable}{lc}
\tablewidth{0pt}
\tablecaption{Parameters of the VLA observations for NGC 24 \label{vla-n24}}
\startdata
\hline\hline
Date of observation\dotfill& 1992 June 14\\
Time on source\dotfill& $5^{\rm h}$\\
Field center (1950.0)\dotfill& $00^{\rm h}07^{\rm m}24.0^{\rm s}$\\
{}& $-25^\circ 14'30.0''$\\
Primary beam at half-power (FWHM)\dotfill& ~32$'$\\
FWHM of synthesised beam\dotfill& $40''\times 40''$\\
Total bandwidth\dotfill& $1.56~{\rm MHz}$\\
Central velocity (heliocentric)\dotfill& $550~{\rm km\,s^{-1}}$\\
Central frequency\dotfill& $1417.2489~{\rm MHz}$\\
Number of channels\dotfill& 128\\
Channel separation\dotfill& $12.2~{\rm kHz}$\\
Velocity Resolution\dotfill& $2.5~{\rm km\,s^{-1}}$\\
RMS noise in channel maps\dotfill& $2.4~{\rm mJy/beam}$\\
Conversion factor, equivalent $1\,$mJy/beam area (low resolution)\\
\hspace{3cm} $40''\times 40''$\dotfill& $0.38~{\rm K}$\\
Maps gridding\dotfill& $10''\times 10''$\\
\vspace{-0.05cm}
                                 {}& natural weighting\\
\vspace{-0.05cm}
                                 {}& no taper\\

Flux calibrator\dotfill& 0134+329\\
\enddata
\end{deluxetable}

\clearpage
 

\begin{deluxetable}{lc}
\tablewidth{0pt}
\tablecaption{Parameters of the VLA observations for NGC 45 \label{vla-n45}}
\startdata
\hline\hline
Date of observation\dotfill& 1992 June 13\\
Time on source\dotfill& $5^{\rm h}$\\
Field center (1950.0)\dotfill& $00^{\rm h}11^{\rm m}30.0^{\rm s}$\\
{}& $-23^\circ 27'60.0''$\\
Primary beam at half-power (FWHM)\dotfill& ~32$'$\\
FWHM of synthesised beam\dotfill& $42''\times 42''$\\
Total bandwidth\dotfill& $1.56~{\rm MHz}$\\
Central velocity (heliocentric)\dotfill& $470~{\rm km\,s^{-1}}$\\
Central frequency\dotfill& $1418.3115~{\rm MHz}$\\
Number of channels\dotfill& 128\\
Channel separation\dotfill& $12.2~{\rm kHz}$\\
Velocity Resolution\dotfill& $2.5~{\rm km\,s^{-1}}$\\
RMS noise in channel maps\dotfill& $1.6~{\rm mJy/beam}$\\
Conversion factor, equivalent $1\,$mJy/beam area (low resolution)\\
\hspace{3cm} $42''\times 42''$\dotfill& $0.34~{\rm K}$\\
Maps gridding\dotfill& $10''\times 10''$\\
\vspace{-0.05cm}
                                 {}& natural weighting\\
\vspace{-0.05cm}
                                 {}& no taper\\                        

Flux calibrator\dotfill& 0134+329\\
\enddata
\end{deluxetable}

\clearpage


\begin{table}
\begin{center}
\caption{$I-$band luminosity profile of NGC 24\label{Iprof-n24}}
\begin{tabular}{cc|cc}
\tableline\tableline
 Radius  &  $\mu_I$  &  Radius  &  $\mu_I$ \\
 (arcsec)  &  (${\rm mag~arcsec^{-2}}$)  &  (arcsec)  &  (${\rm mag~arcsec^{-2}}$)
 \\
\tableline
1.96& 18.54  & 23.37& 19.42\\
2.37& 18.57  & 28.27& 19.48\\
2.87& 18.59  & 34.21& 19.61\\
3.48& 18.61  & 41.39& 19.87\\
4.21& 18.63  & 50.09& 20.14\\
5.09& 18.64  & 60.60& 20.58\\
6.15& 18.69  & 73.33& 21.11\\
7.45& 18.83  & 88.73& 21.38\\
9.01& 18.99  & 107.4& 21.95\\
10.90& 19.11 & 129.9& 22.41\\
13.19& 19.19 & 157.2& 23.11\\
15.96& 19.26 & 190.2& 24.22\\
19.31& 19.33 & {}&{}\\
\tableline
\end{tabular}
\end{center}
\end{table}


\begin{table}
\begin{center}
\caption{$B-$band Luminosity profile of NGC 45 (Romanishin et al. 1983)\label{Bprof-n45}}
\begin{tabular}{cc|cc}
\tableline\tableline
 Radius  &  $\mu_B$  &  Radius  &  $\mu_B$ \\
 (arcsec)  &  (${\rm mag~arcsec^{-2}}$)  &  (arcsec)  &  (${\rm mag~arcsec^{-2}}$)
 \\
\tableline
 2.3 & 22.17 &  142.8& 24.95\\
11.7 & 22.43 &  156.8& 25.06\\
16.4 & 22.54 &  170.9& 25.16\\
21.1 & 22.66 &  184.9& 25.36\\
25.8 & 22.77 &  199.0& 25.64\\
30.4 & 22.90 &  213.0& 25.89\\
35.1 & 23.08 &  227.1& 25.99\\
44.5 & 23.22 &  241.1& 26.38\\
58.5 & 23.53 &  255.1& 26.96\\
72.6 & 23.79 &  269.2& 27.12\\
86.6 & 24.12 &  283.2& 27.70\\
100.7& 24.29 &  297.3& 27.99\\
114.7& 24.49 &  311.3& 28.43\\
128.8& 24.73 &  325.4& 29.02\\
\tableline
\end{tabular}
\end{center}
\end{table}

\clearpage


\begin{deluxetable}{lc}
\tablewidth{0pt}
\tablecaption{Optical parameters of NGC 24 \label{opar-n24}}
\startdata
\hline\hline
R.A. (2000) $^{\rm a}$\dotfill&  $00^{\rm h}09^{\rm m}56.6^{\rm s}$\\
DEC. (2000) $^{\rm a}$\dotfill&  $-24^\circ57'43.0''$\\
Type $^{\rm a}$\dotfill&  SA(s)c III\\
Distance (Mpc) $^{\rm b}$\dotfill& 6.8\\
&(1$'$ = 2.0 kpc)\\
Mean axis ratio $^{\rm c}$ (q = b/a)\dotfill& ${\rm q}= 0.26$\\
Inclination $^{\rm c}$ $({\rm q_0} = 0.15)^{\rm d}$\dotfill& ${\rm i}= 78^\circ$\\
Mean position angle $^{\rm c}$\dotfill& $\theta=  225^\circ $\\
Parameters at $\mu_B = 25.0\,{\rm mag\,arcsec^{-2}}$:\\
\hspace{1cm} Major axis diameter\dotfill& ${\rm D_{25}} = 5.9'$\\
\hspace{1cm} Minor axis diameter \dotfill& ${\rm d_{25}} = 1.6'$\\
Holmberg radius ($\mu_B = 26.6\,{\rm mag\,arcsec^{-2}}$)\dotfill& ${\rm R_{Ho}}= 4.0' $\\
Exponential disk parameters:\\
\hspace{1cm }Corrected central surface brightness $^{\rm e}$\dotfill& ${\rm I(0)_c} = 20.67$\\
{}&${\rm B(0)_c} = 22.12$\\
\hspace{1cm }Scale length (kpc)\dotfill& $\alpha^{-1} = 1.42 $\\
Total apparent B magnitude\dotfill& ${\rm B_T} = 12.13$\\
Corrected apparent B magnitude\dotfill& ${\rm B_T^{0,i}}= 11.75$\\
Corrected absolute B magnitude\dotfill& ${\rm M_B^{0,i}} = -17.41$\\
Total blue luminosity $^{\rm f}$\dotfill& ${\rm L_T(B) =  1.4\times10^9\,L_{\odot}}$\\
\enddata
\tablenotetext{a}{de Vaucouleurs  et al. 1991 (RC3).}
\tablenotetext{b}{Tully 1988.}
\tablenotetext{c}{For $1.5' \le  R \le 3.0'$.}
\tablenotetext{d}{$\cos ^2 i = \frac{q^2 - q_0^2}{1-q_0^2}$ following
Bottinelli et al 1983.}
\tablenotetext{e}{Line-of-sight integration correction = $2.5\log R_{25} =
1.575~({\rm RC3})$.}
\tablenotetext{f}{ With $M_{B\odot} = +5.43$ (Allen 1976).}
\end{deluxetable}

\clearpage


\begin{deluxetable}{lc}
\tablewidth{0pt}
\tablecaption{Optical parameters of NGC 45 \label{opar-n45}}
\startdata
\tableline\tableline
R.A. (2000) $^{\rm a}$\dotfill&  $00^{\rm h}14^{\rm m}03.2^{\rm s}$\\
DEC. (2000) $^{\rm a}$\dotfill&  $-23^\circ11'01.0''$\\
Type $^{\rm a}$\dotfill&  SA(s)dm IV-V\\
Distance (Mpc) $^{\rm b}$\dotfill& 5.9 \\
& (1$'$ = 1.72 kpc)\\
Mean axis ratio $^{\rm d}$ (q = b/a)\dotfill& ${\rm q}= 0.73$\\
Inclination $^{\rm c}$ $({\rm q_0} = 0.22)^{\rm e}$\dotfill& ${\rm i} = 55^\circ \pm 5^\circ$\\
Position angle $^{\rm c}$\dotfill& $\theta = 145^\circ \pm 5^\circ  $\\
Parameters$^{\rm d}$ at $\mu_B = 25.0\,{\rm mag\,arcsec^{-2}}$:\\
\hspace{1cm} Major axis diameter\dotfill& ${\rm D_{25}} = 5.8'$\\
\hspace{1cm} Minor axis diameter \dotfill& ${\rm d_{25}} = 4.2'$\\
Holmberg radius ($\mu_B = 26.6\,{\rm mag\,arcsec^{-2}}$)$^{\rm d}$\dotfill& ${\rm R_{Ho}} = 4.8'$\\
Exponential disk parameters$^{\rm d}$:\\
\hspace{1cm }Corrected central surface brightness \dotfill&${\rm B(0)_c} = 22.51$\\
\hspace{1cm }Scale length (kpc)\dotfill& $\alpha^{-1} = 2.20$\\
Total apparent B magnitude$^{\rm d}$\dotfill& ${\rm B_T} = 11.48$\\
Corrected apparent B magnitude\dotfill& ${\rm B_T^{0,i}}= 11.40$\\
Corrected absolute B magnitude\dotfill& ${\rm M_B^{0,i}} = -17.45$\\
Total blue luminosity $^{\rm f}$\dotfill& ${\rm L_T(B) = 1.4\times10^9\,L_{\odot}}$\\
\enddata
\tablenotetext{a}{de Vaucouleurs et al. 1991 (RC3).}
\tablenotetext{b}{Tully 1988.}
\tablenotetext{c}{Puche \& Carignan 1988.}
\tablenotetext{d}{Romanishin et al. 1983.}
\tablenotetext{e}{$\cos ^2 i = \frac{q^2 - q_0^2}{1-q_0^2}$ 
following Bottinelli et al 1983.}
\tablenotetext{f}{With $M_{B\odot} = +5.43$ (Allen 1976).}
\end{deluxetable} 

\clearpage


\begin{table}
\begin{center}
\caption{\hi Rotation Curve for NGC 24 \label{rc-n24}}
\begin{tabular}{ccc|ccc}
\tableline\tableline
 Radius  &  $V_{rot}$  & Error &  Radius &  $V_{rot}$ &  Error \\
 (arcsec)  &  (\kms) & (\kms)  &  (arcsec)  &  (\kms)  &  (\kms) \\
\tableline
 40   &   56.6   &   6.4   & 200   &   103.6  &   2.2    \\
 80   &   82.5   &   2.7   & 240   &   106.2  &   1.8  \\
 120   &   91.8   &   0.8  & 280   &   109.4  &   3.5   \\
 160   &   99.2   &   0.8  & 320   &   109.7  &   2.0  \\
\tableline
\end{tabular}
\end{center}
\end{table}


\begin{table}
\centering
\caption{\hi Rotation Curve and Position Angle of the kinematical major axis for NGC 45 \label{rc-n45}}
\begin{tabular}{ccccc|ccccc}
\tableline\tableline
 Radius  &  $V_{rot}$  & Error &  $P.A.$  & Error &  Radius &  $V_{rot}$ &  Error &  $P.A.$  & Error \\
 (arcsec)  &  (\kms) & (\kms)  &  ($\degr$)  &  ($\degr$) &  (arcsec)  &  (\kms)  &  (\kms)  &  ($\degr$)  &  ($\degr$) \\
\tableline
42   &53.2  &  3.6  & 144.0   &  11.0  &  336  & 105.9 &  0.3  & 139.3   & 0.2 \\
84   &70.1  &  1.3  & 148.6   &  5.9   &  378  & 103.4 &  1.4  & 136.5   & 0.1 \\
126  &79.1  &  1.7  & 148.2   &  1.0   &  420  & 101.0 &  0.9  & 132.7   & 1.6 \\
168  &87.4  &  0.3  & 147.0   &  0.3   &  462  & 100.9 &  1.0  & 129.0   & 1.5 \\
210  &93.4  &  0.4  & 147.2   &  0.7   &  504  & 99.9  &  1.5  & 126.4   & 1.3 \\
252  &97.7  &  2.2  & 145.9   &  1.3   &  546  & 99.8  &  0.3  & 124.6   & 0.7 \\
294  &102.3 &  1.6  & 143.0   &  1.0   &  588  & 100.0 &  0.6  & 124.9   & 0.4 \\    
\tableline
\end{tabular}
\end{table}

\clearpage
 

\begin{deluxetable}{lc}
\tablewidth{0pt}
\tablecaption{``Best-Fit'' Two Component Model for NGC 24 \label{mod-n24}}
\startdata
\tableline\tableline
{\it Luminous disk component:}\\
\rml &                                 2.5$~{\cal M}_ \odot /L_ \odot$\\
${\cal M}_{\rm disk}$&                      $3.4\times 10^9~{\cal M}_ \odot$\\
${\cal M}_{\rm HI+He}$&                        $7.4\times 10^8~{\cal M}_ \odot$\\
\\

{\it Dark halo component:}\\
${\rm r_c}$&                          5.6~kpc\\
$\sigma$&                             65~${\rm km\,s^{-1}}$\\
$\rho_0$&                             0.022~${\cal M}_ \odot \,{\rm pc^{-3}}$\\
\\

{\it At $ R_{HO}$ (r=7.9\,kpc) :}\\
${\rm \rho_{halo}}$&                    0.004~${\cal M}_ \odot \,{\rm pc^{-3}}
$\\
${\cal M}_{\rm dark}/{\cal M}_{\rm lum}$&    4.0\\
$({\cal M}/\rm L_B)_{dyn}$&                     15.3$~{\cal M}_\odot /L_\odot$\\
${\cal M}_{\rm dark+lum}$&                   $2.0\times 10^{10}~{\cal M}_ \odot$
\\
\\

{\it At last measured point (R=10.5\,kpc):}\\
${\rm \rho_{halo}}$&                        0.0022~${\cal M}_ \odot
\,{\rm pc^{-3}}$\\
${\cal M}_{\rm dark}/{\cal M}_{\rm lum}$&        6.0\\
$({\cal M}/\rm L_B)_{dyn}$&                      21.2$~{\cal M}_ \odot /L_ \odot
$\\
${\cal M}_{\rm dark+lum}$&                  $2.8\times 10^{10}~{\cal M}_ \odot$
\\
\enddata
\end{deluxetable}


\begin{deluxetable}{lc}
\tablewidth{0pt}
\tablecaption{``Best-Fit'' Two Component Model for NGC 45 \label{mod-n45}}
\startdata
\tableline\tableline
{\it Luminous disk component:}\\
\rml &                                 5.2$~{\cal M}_ \odot /L_ \odot$\\
${\cal M}_{\rm disk}$&                      $5.3\times 10^9~{\cal M}_ \odot$\\
${\cal M}_{\rm HI+He}$&                        $2.1\times 10^9~{\cal M}_ \odot$\\
\\

{\it Dark halo component:}\\
${\rm r_c}$&                          6.2~kpc\\
$\sigma$&                             55~${\rm km\,s^{-1}}$\\
$\rho_0$&                             0.013~${\cal M}_ \odot \,{\rm pc^{-3}}$\\
\\

{\it At $ R_{HO}$ (r=8.3\,kpc) :}\\
${\rm \rho_{halo}}$&                   0.003~${\cal M}_ \odot \,{\rm pc^{-3}}$
\\
${\cal M}_{\rm dark}/{\cal M}_{\rm lum}$&               1.8\\
$({\cal M}/\rm L_B)_{dyn}$&                  18.0$~{\cal M}_ \odot /L_ \odot$\\
${\cal M}_{\rm dark+lum}$&                   $1.8\times 10^{10}~{\cal M}_ \odot$
\\
\\

{\it At last measured point (R=16.7\,kpc):}\\
${\rm \rho_{halo}}$&                     0.0005~${\cal M}_ \odot \,{\rm pc^{-3}
}$\\
${\cal M}_{\rm dark}/{\cal M}_{\rm lum}$&               4.0\\
$({\cal M}/\rm L_B)_{dyn}$&                  36.2$~{\cal M}_ \odot /L_ \odot$\\
${\cal M}_{\rm dark+lum}$&                   $3.7\times 10^{10}~{\cal M}_ \odot$\\
\enddata
\end{deluxetable}


\begin{deluxetable}{lc}
\tablewidth{0pt}
\tablecaption{Two Component Model for NGC 45 with (M/L) = 2.0 \label{mod-n45color}}
\startdata
\tableline\tableline
{\it Luminous disk component:}\\
\rml &                                 2.0$~{\cal M}_ \odot /L_ \odot$\\
${\cal M}_{\rm disk}$&                      $2.0\times 10^9~{\cal M}_ \odot$\\
${\cal M}_{\rm HI+He}$&                        $2.1\times 10^9~{\cal M}_ \odot$\\
\\

{\it Dark halo component:}\\
${\rm r_c}$&                          4.2~kpc\\
$\sigma$&                             58~${\rm km\,s^{-1}}$\\
$\rho_0$&                             0.032~${\cal M}_ \odot \,{\rm pc^{-3}}$\\
\\

{\it At $ R_{HO}$ (r=8.3\,kpc) :}\\
${\rm \rho_{halo}}$&                   0.003~${\cal M}_ \odot \,{\rm pc^{-3}}$
\\
${\cal M}_{\rm dark}/{\cal M}_{\rm lum}$&               4.7\\
$({\cal M}/\rm L_B)_{dyn}$&                  18.8$~{\cal M}_ \odot /L_ \odot$\\
${\cal M}_{\rm dark+lum}$&                   $1.9\times 10^{10}~{\cal M}_ \odot$
\\
\\

{\it At last measured point (R=16.7\,kpc):}\\
${\rm \rho_{halo}}$&                     0.0005~${\cal M}_ \odot \,{\rm pc^{-3}
}$\\
${\cal M}_{\rm dark}/{\cal M}_{\rm lum}$&               7.8\\
$({\cal M}/\rm L_B)_{dyn}$&                  35.8$~{\cal M}_ \odot /L_ \odot$\\
${\cal M}_{\rm dark+lum}$&                   $3.7\times 10^{10}~{\cal M}_ \odot$\\
\enddata
\end{deluxetable}

\end{document}